\documentclass[%
 aip,
 amsmath,amssymb,
 preprint,%
]{revtex4-1}

\usepackage{graphicx}
\usepackage{dcolumn}
\usepackage{bm}

\usepackage[utf8]{inputenc}
\usepackage[T1]{fontenc}
\usepackage{mathptmx}

\begin{document}


\title[Ab Initio Bethe-Salpeter Equation Approach to Neutral Excitations in Molecules]{Ab Initio Bethe-Salpeter Equation Approach to Neutral Excitations in Molecules with Numeric Atom-Centered Orbitals}

\author{Chi Liu}
\affiliation
{Department of Chemistry, Duke University, Durham, North Carolina 27708, United States}
\author{Jan Kloppenburg}
\affiliation{Institute of Condensed Matter and Nanoscience, Universit\'e Catholique de Louvain, Louvain-la-Neuve, 1348, Belgium}
\author{Xinguo Ren}
\affiliation{CAS Key Laboratory of Quantum Information, University of Science and Technology of China, Hefei, Anhui 230026, China}
\author{Heiko Appel}
\affiliation{Max Planck Institute for the Structure and Dynamics of Matter, Center for Free Electron Laser Science, 22761 Hamburg, Germany}
\author{Yosuke Kanai}
\affiliation
{Department of Chemistry, University of North Carolina, Chapel Hill, North Carolina 27599, United States}
\author{Volker Blum}
\email
{volker.blum@duke.edu}
\affiliation
{Department of Mechanical Engineering and Materials Science, Duke University, Durham, North Carolina 27708, United States}
\altaffiliation
{Department of Chemistry, Duke University, Durham, North Carolina 27708, United States}

\date{\today}

\begin{abstract}
The Bethe-Salpeter equation (BSE) based on $GW$ quasiparticle levels is a successful approach for calculating the optical gaps and spectra of solids and also for predicting the neutral excitations of small molecules. We here present an all-electron implementation of the $GW$+BSE formalism for molecules, using numeric atom-centered orbital (NAO) basis sets. We present benchmarks for low-lying excitation energies for a set of small organic molecules, denoted in the literature as ``Thiel's set''. Literature reference data based on Gaussian-type orbitals are reproduced to about one meV precision for the molecular benchmark set, when using the same $GW$ quasiparticle energies and basis sets as the input to the BSE calculations. For valence correlation consistent NAO basis sets, as well as for standard NAO basis sets for ground state density-functional theory with extended augmentation functions, we demonstrate excellent convergence of the predicted low-lying excitations to the complete basis set limit. A simple and affordable augmented NAO basis set denoted ``tier2+aug2'' is recommended as a particularly efficient formulation for production calculations. We finally demonstrate that the same convergence properties also apply to linear-response time-dependent density functional theory within the NAO formalism. 
\end{abstract}

\maketitle

\section{\label{sec:introduction}Introduction}
Predicting the neutral (including optical) excitations of molecules and materials is of fundamental importance in photovoltaics, optoelectronics, and other technologically relevant areas. Several distinct types of computational formalisms are frequently employed in the community for this purpose, including: wavefunction-based methods, e.g., equation-of-motion coupled cluster (EOM-CC)\cite{eom_cc1,eom_cc2,eom_cc3} or complete active space second-order perturbation theory (CASPT2),\cite{ROOS1980157,caspt2_1,caspt2_2,caspt2_3} the quantum Monte Carlo method,\cite{qmc1,qmc2,qmc3,qmc4} linear-response time-dependent density functional theory (LR-TDDFT)\cite{tddft1, tddft2, casida, casida2} or the Bethe-Salpeter equation (BSE) approach in the context of many-body perturbation theory (MBPT).\cite{hedin, bse1951, louie86prb, louie1998prl, louie2000prb, Strinati1988, reining2002review, Blase2018review} EOM-CC and CASPT2 have been shown to produce highly accurate values for small and mid-sized molecules, when combined with sufficiently high-quality basis sets. They are therefore often used as a trusted reference, \cite{eom_cc1,eom_cc2,eom_cc3,caspt2_1,caspt2_2,caspt2_3,thiel} although their applicability to larger systems is somewhat limited by the associated computational cost. LR-TDDFT has been widely applied to predict optical excitations for molecules due to its computational efficiency and often reasonable accuracy, especially when combined with carefully designed exchange-correlation (XC) functionals.\cite{tddft_functional_review, thielTddft, functional_BmLBLYP, BNL, OT-RSH, CAM-B3LYP}. However, LR-TDDFT calculations can encounter problems for charge-transfer (CT) excitations\cite{ct1, ct2}, especially when used with a simple XC functional such as the adiabatic local density approximation (LDA)\cite{ks1965} and generalized gradient approximations (GGAs).\cite{pbe} In LR-TDDFT, including long-range exact exchange in the XC functional can mitigate this problem.\cite{solve_ct_2017, BNL, OT-RSH, CAM-B3LYP} 

The BSE approach is founded upon MBPT, based on Green's function ($G$) theory and the idea of using the screened Coulomb interaction $W$.\cite{hedin} The BSE formalism was originally proposed in the field of nuclear physics in 1950s.\cite{bse1951} Combined with the $GW$ approximation in MBPT,\cite{hedin} the BSE approach has been shown to successfully approximate the optical spectra of solids\cite{louie1998prl, ReiningPrl1998, BenedictPrl1998, ShirleyPrl1998, louiePrl2004} and later work demonstrates similar applicability to excitations in atoms and molecules.\cite{louie2000prb, tiago2006bsePrb, bseTdaPrb2009, bseWaterBlase, MolGWBse, bseBenchBlase, isBseAccurateBlase, bse_TURBOMOLE, bseCationDye2017, excitonTypeBse, hePrl2017, tdaBseTriplet2017} The $GW$ approach\cite{hedin,gwAulbur1999,Strinati1988} allows one to predict fundamental gaps -- i.e., gaps between highest occupied molecular orbital (HOMO) and lowest unoccupied molecular orbital (LUMO) quasiparticle states -- as well as single-quasiparticle excitation spectra that are more accurate than those obtained by standard density functional theory (DFT) for a wide range of systems, including both solids and molecules.\cite{gwNiOPrl1995, gwBandPrl2004,Kaplan2015,Baumeier2014,Caruso2012Prb,Marom2012Prb,Bruneval2012,Galli2013,Caruso2013,Brunevaljctc2013,Faber2011prb,vanSetten2013jctc,Faber2014,Atallaprb2013,RenRI2012} The description of optical excitations within the BSE approach then uses charged excitations, i.e., electron removal and addition excitations, from the $GW$ approach as its input. The BSE method based on the $GW$ method has several formal advantages over LR-TDDFT. The electron-hole interaction in the BSE approach has the correct asymptotic behavior for both solids and molecules, which is not captured by LR-TDDFT formalism without long range exchange component.\cite{MolGWBse} CT excitations, which are problematic for LR-TDDFT especially with LDA and GGA functionals, can be efficiently and accurately predicted by the BSE approach.\cite{blaseBseCT2011,galliBseCT2010,bseCtBaumeier2012,excitonTypeBse} This has been demonstrated for CT excitations including both intermolecular and intramolecular types in systems such as simple dipeptides,\cite{blaseDipeptide2013, galliBseCT2010} and more complex fullerene/polymer aggregates.\cite{bseCtBaumeier2012,faberPrb2012} 

Calculations of BSE excitation energies within MBPT usually adopt a three-step procedure: (i) Evaluate the Kohn-Sham (KS) \cite{ks1965} or generalized Kohn-Sham (gKS) \cite{gks} DFT orbitals. (ii) Apply self-energy corrections at the $G^0W^0$ level or self-consistent $GW$ level ($G^0W^0$@DFT or $GW$@DFT; $G^0$ stands for the Green's function of a non-interacting reference system and $W^0$ is the screened Coulomb interaction of that reference system).\cite{hedin,gwAulbur1999,Strinati1988,louie2000prb} (iii) Solve the BSE (in practice, an approximate version thereof, see below) based on the $G^0W^0$ or $GW$ quasiparticle energies and screened and unscreened Coulomb integrals of (g)KS orbitals (BSE@$G^0W^0$@DFT or BSE@$GW$@DFT).\cite{bse1951,bseBenchBlase,MolGWBse,bse_TURBOMOLE} BSE implementations exist in different computational packages based on different basis functions, e.g., MolGW\cite{MolGW} and Turbomole,\cite{turbmole,bse_TURBOMOLE} which are based on Gaussian-type orbital (GTO) basis sets, or BerkeleyGW,\cite{berkeleygw} Yambo,\cite{yambo} Exciting,\cite{exciting} ABINIT,\cite{abinit1,abinit2} VASP\cite{vasp1996} and Quantum Espresso,\cite{quantumEspresso} which are based on plane waves.

The present work introduces an accurate implementation of the BSE formalism utilizing compact and efficient numeric atom-centered orbital (NAO) basis sets\cite{Blum2009,Igor2015NJP} in the context of the all-electron electronic structure code FHI-aims.\cite{Blum2009, Havu2009, RenRI2012, IhrigNJP2015} To obtain the two-electron Coulomb and screened Coulomb interaction matrix elements, we use an efficient and highly accurate variant of the resolution-of-identity (RI) technique.\cite{RenRI2012} In FHI-aims, this RI technique is the numerical foundation for all methods beyond semilocal DFT, including Hartree-Fock (HF), hybrid density functionals, the random-phase approximation (RPA), second-order M{\o}ller-Plesset perturbation theory (MP2) and the $GW$ method.\cite{Ren2012NJP,IhrigNJP2015} Our current implementation also uses the ELSI infrastructure\cite{YU2017} and the ELPA eigenvalue solver\cite{elpa2014} for parallel eigenvalue solutions. 

The paper is organized as follows. In Section \ref{sec:methods}, we introduce the $GW$+BSE formalism in the context of MBPT. In Section \ref{sec:implementation}, we discuss the details of our implementation. In Section \ref{sec:results}, we demonstrate the numerical correctness of our BSE implementation by comparing the excitation energies computed by FHI-aims and by the MolGW code\cite{MolGWBse, MolGW} with GTO correlation-consistent basis sets for Thiel's molecular benchmark set.\cite{thiel, thielTddft} In assessing our BSE implementation, we emphasize the dependence of the BSE results on the $GW$ quasiparticle energies. We then study the convergence behavior of excitation energies to the complete basis set limit, combining standard NAO basis sets  for ground state DFT (FHI-aims-2009)\cite{Blum2009} or valence correlation consistent NAO basis sets (NAO-VCC-nZ)\cite{Igor2015NJP} with extended augmentation functions (NAO+aug)  We demonstrate that the standard FHI-aims-2009 basis sets give essentially basis set converged numerical results for low-lying optical excitation energies when combined with a few extended augmentation functions (NAO+aug basis sets) that are also commonly included in Gaussian-type basis sets. Finally, similar convergence behavior is demonstrated for LR-TDDFT with adiabatic LDA as the kernel.\cite{ks1965,pwlda} 

\section{\label{sec:methods}Methods}
Typical calculations of the neutral (optical) excitation energies of molecules using the BSE approach adopt the following three-step procedure, which is utilized by a wide range of electronic structure packages for calculations of neutral (optical) excitation properties in the framework of MBPT.\cite{vasp1996,berkeleygw,abinit2,exciting,yambo,MolGWBse,bseBenchBlase,bse_TURBOMOLE} 

(i) The initial step is performed by solving the self-consistent (g)KS equations with an approximate functional for the exchange-correlation energy $E_{\textrm{xc}}$. Common choices for $E_{\textrm{xc}}$ are the LDA, GGAs, HF and hybrid functionals. In KS theory (e.g., LDA and GGAs), we define $\hat{v}_{\textrm{xc}}$ as the functional derivative of $E_{\textrm{xc}}$ with respect to the electron density. In the gKS case (HF and hybrid functionals), $\hat{v}_{\textrm{xc}}$ is the functional derivative with respect to the set of orbitals $\psi_l$. In either case, the $\psi_l$ are constructed as:
\begin{equation} \label{ks}
	\hat{h}^{\textrm{(g)KS}} |\psi_l\rangle  = \epsilon_l |\psi_l\rangle \, ,
\end{equation}
\begin{equation} \label{hamilt}
	\hat{h}^{\textrm{(g)KS}} = \hat{t}_\textrm{s} + \hat{v}_{\textrm{ext}} + \hat{v}_{\textrm{H}} + \hat{v}_{\textrm{xc}} \, .
\end{equation}
Equation (\ref{ks}) states the electronic (g)KS single-particle equations for the effective single-particle orbitals $\psi_l$ and eigenvalues $\epsilon_l(l = 1, 2, ..., N_{orbit})$. Equation (\ref{hamilt}) details the gKS Hamiltonian $\hat{h}^{\textrm{(g)KS}}$, including the effective single-particle kinetic energy (with relativistic corrections) $\hat{t}_\textrm{s}$, the external potential $\hat{v}_{\textrm{ext}}$, the electrostatic or Hartree potential of the electron density $\hat{v}_{\textrm{H}}$ and the exchange-correlation potential $\hat{v}_{\textrm{xc}}$. 
The underlying orbitals $\{\psi_l\}$ are here expanded in a basis of NAOs $\{\varphi_i, i = 1, 2, ..., N_{basis}\}$,
\begin{equation}
\psi_l = \sum_i c_{li} \varphi_i \, ,
\end{equation}
where the NAOs are of the form\cite{Blum2009}
\begin{equation}
\varphi_i (\mathbf{r}) = \frac{u_i(r)}{r} Y_{lm}(\Omega) \, .
\end{equation}
$\mathbf{r}$ is a position vector with respect to the nucleus, $r$ is its modulus, and $\Omega$ the corresponding solid angle. In the FHI-aims code, the $u_i$ are numerically tabulated functions, defined as cubic splines in units of a logarithmic grid. $Y_{lm}$ are the real-valued spherical harmonics, and $l, m$ are implicitly included in the basis function index $i$. The eigenvalues and eigenfunctions produced by the initial step serve as a first guess for the quasiparticles and are used to evaluate the Coulomb interaction, the screened Coulomb interaction and the $GW$ self-energy in the subsequent $GW$ and BSE@$GW$ steps. Although, in the non-periodic case, $\psi_l$ can be chosen to be real-valued, we include complex conjugates in the derivations below.

(ii) A perturbative $GW$ approach is then applied to obtain the quasiparticle energies\cite{RenRI2012}
\begin{equation} \label{gw}
	\epsilon^{GW}_{l} = \epsilon_{l} + \langle \psi_{l} | \Sigma^{GW}(\epsilon^{GW}_{l}) - \hat{v}_{\textrm{xc}} |\psi_{l}\rangle \, ,
\end{equation}
where $\epsilon^{GW}_{l}$ is the quasiparticle energy. By convention, the arguments $\epsilon^{GW}_{l}$ used to evaluate the self-energy $\Sigma^{GW}$ on the right-hand side are updated self-consistently until they match the $\epsilon^{GW}_{l}$ values obtained on the left-hand side, even though the function $\Sigma^{GW}(\omega)$ itself is not further updated in the process. The $GW$ self-energy is calculated from the Green's function $G$ and the screened Coulomb potential $W$ following the $GW$ approximation proposed by Hedin\cite{hedin}:
\begin{equation} \label{selfenergy}
	\Sigma^{GW}(\textbf{r}, \textbf{r}^\prime, \omega) = \frac{i}{2\pi} \int d\omega^\prime G(\textbf{r}, \textbf{r}^\prime, \omega + \omega^\prime) W(\textbf{r}, \textbf{r}^\prime, \omega^\prime) e^{i\omega^\prime \eta} \, .
\end{equation}
In the single-shot perturbative $GW$ (i.e., $G^0W^0$) approach, the Green's function $G$ is approximated by the non-interacting Green's function $G^0$, which is calculated from single-particle orbitals $\psi_l$ and orbital energies $\epsilon_l$:\cite{RenRI2012}
\begin{equation} \label{gwg}
G^0(\textbf{r}, \textbf{r}^\prime, \omega) = \sum_l \frac{\psi_l(\textbf{r}) \psi_l^\dagger(\textbf{r}^\prime)}{\omega - \epsilon_l - \textrm{i}\eta \textrm{sgn}(\epsilon_\textrm{F} - \epsilon_l)} \, .
\end{equation}
$\epsilon_\textrm{F}$ is the Fermi energy and $\eta$ is a positive infinitesimal. The screened Coulomb potential $W^0$ is calculated from the dielectric function $\varepsilon$ as\cite{RenRI2012}
\begin{equation}
W^0(\textbf{r}, \textbf{r}^\prime, \omega^\prime) = \int d\textbf{r}^{\prime\prime} \varepsilon^{-1}(\textbf{r}, \textbf{r}^{\prime\prime}, \omega^\prime) v(\textbf{r}^{\prime\prime}, \textbf{r}^\prime) \, ,
\end{equation}
where the dielectric function $\varepsilon$ is obtained at the RPA level, using DFT results.
The $G^0W^0$ self-energy can be calculated using an exact analytic treatment on the real axis, which is the case in the MolGW package.\cite{MolGW,gw_exact_freq} We refer to Section III.C of Ref. \citenum{gw_exact_freq} for the details of this formalism. This treatment is limited to small systems. Instead, two-pole\cite{godbyneeds1995} and Pad\'{e}\cite{pades} approximations are implemented in the FHI-aims code for the evaluation of the self-energy on the real axis.\cite{RenRI2012}
Both of these approximations are based on an exact treatment of $G^0$, $W^0$, and the self-energy $\Sigma^{G^0W^0}$ on the imaginary frequency axis. $\Sigma^{G^0W^0}$ is then extended to the real axis by performing an analytic fit of the data on the imaginary axis to a function with a form that has poles on the real axis. This process is usually referred to as ``analytic continuation". The smooth behavior of all quantities ($G^0$, $W^0$, $\Sigma^{G^0W^0}$) on the imaginary frequency axis significantly reduces the number of frequency points needed, compared to a full frequency integration along the real frequency axis.\cite{gw100} 
Specifically, the self-energy is approximated to have the following mathematical form in the complex plane in the two-pole approximation:\cite{godbyneeds1995}
\begin{equation}\label{2pole}
\Sigma_{ij}(z) \approx \sum^{2}_{n=1} \frac{a_n}{z + b_n} \, ,
\end{equation}
where the values of $a_n$ and $b_n$ depend on the indices $i$ and $j$. 
In the Pad\'{e} approximation, the self-energy is expressed as\cite{pades}
\begin{equation}\label{pade}
\Sigma_{ij}(z) \approx \frac{a_0 + a_1 z + ... + a_{(N-2)/2}(z)^{(N-2)/2}}{1 + b_1 z + ... + b_{N/2}(z)^{N/2}} \, ,
\end{equation}
where N denotes the total number of parameters in the Pad\'{e} approximation. We note already here that the Pad\'{e} approximation can be more accurate than the two-pole approximation to represent the true self-energy, but that the Pad\'{e} approximation is also, in practice, more prone to numerical problems, including non-unique solutions that can be difficult to control without manual inspection of all resulting eigenvalues. In addition to the two approaches mentioned above, another, more elaborate approach to evaluate the self-energy directly on the real axis by contour deformation (CD) was implemented in FHI-aims by Golze and coworkers\cite{GolzeGW} while this paper was being completed. We do not assess this approach here because our emphasis here is on the BSE but we note that essentially exact $G^0W^0$ input data to the BSE are expected from the CD approach. On the other hand, the analytical continuation of $\Sigma$ according to Eqs. (\ref{2pole}) or (\ref{pade}) is advantageous over the CD approach in terms of computational cost, both in terms of the base cost (often called the prefactor) and in terms of the scaling exponent with system size if the number of needed $G^0W^0$ eigenvalues scales with the size of the system.\cite{GolzeGW}

Here we perform one-shot perturbative $G^0W^0$ calculations based on a fixed DFT or HF reference. The quasiparticle energy in equation (\ref{gw}) is thus rewritten as $\epsilon^{G^0W^0}_{l}$. Some studies investigate the effect of iterating the $GW$ equations by updating the eigenvalues in equation (\ref{gwg}) by those calculated from equation (\ref{gw}), whereas the wavefunctions $\psi_l$ are kept at the DFT level.\cite{bseBenchBlase,blase_prl_he,blase_sc_gw_bse2016} This procedure, denoted as eigenvalue self-consistent $GW$, is reported to give better agreement with experimental results and wavefunction-based reference methods compared with single-shot $G^0W^0$ for some systems.\cite{blase_sc_gw_bse2016,blase_prl_he,bseBenchBlase,isBseAccurateBlase}

(iii) The BSE is a Dyson-like equation for the two-particle correlation function $L$:\cite{louie2000prb,Strinati1988,MolGWBse}
\begin{equation}
    L(12;1'2') = L_0(12;1'2') + \int d(3456) L_0(14;1'3) K(35;46) L(62;52') \, ,
\end{equation}
where the set of variables 1, 2, etc. are short for position, time, and spin $(\mathbf{r}_1,t_1,\sigma_1)$, $(\mathbf{r}_2,t_2,\sigma_2)$, etc. $L(12;1'2')$ is the electron-hole correlation function which describes the probability amplitude of an electron propagating from $1^\prime$ to 2 and a hole propagating from 1 to $2^\prime$.\cite{Strinati1988} $L_0(12;1'2')$ represents the correlation function of the non-interacting system as defined below in equation (\ref{correlation0}). $K(35;46)$, usually called the electron-hole interaction kernel, is the screened interaction between the electron and the hole (including bare exchange). $L_0$ and $K$ can be expressed in the following equations:\cite{louie2000prb}
\begin{equation} \label{correlation0}
    L_0(12;1'2') = G^0(1, 2') G^0(2, 1') \, ,
\end{equation}
\begin{equation} \label{kernel}
    K(35;46) = \frac{\delta M(3, 4)}{\delta G(6, 5)} \, ,
\end{equation}
where $G$ is the one-particle Green's function and $M$ is equal to the sum of the self-energy and the Hartree potential:
\begin{equation} \label{hartree_self}
    M(3, 4) = v_\mathrm{H}(3)\delta(3, 4) + \Sigma(3, 4) \, .
\end{equation}
By applying equation (\ref{hartree_self}) to equation (\ref{kernel}), performing a time-energy Fourier transformation and ignoring the dynamical properties of $W$,\cite{louie2000prb} the BSE kernel can be simplified to:
\begin{equation} \label{kernel_static}
    K(\mathbf{r}_3, \mathbf{r}_5, \mathbf{r}_4, \mathbf{r}_6) = -iv(\mathbf{r}_3 - \mathbf{r}_5) \delta(\mathbf{r}_3 - \mathbf{r}_4) \delta(\mathbf{r}_5 - \mathbf{r}_6) + iW(\mathbf{r}_3, \mathbf{r}_4, \omega = 0) \delta(\mathbf{r}_3 - \mathbf{r}_5) \delta(\mathbf{r}_4 - \mathbf{r}_6) \, ,
\end{equation}
where the variables 3, 4, etc. are reduced to $\mathbf{r}_3, \mathbf{r}_4$, etc. $v$ is the bare Coulomb interaction. $W$ is the screened Coulomb interaction, with the frequency-dependence ignored.\cite{louie2000prb} This approximation means that the actual BSE part (once the $GW$ quasiparticle energies are fixed) is independent of a particular analytical continuation choice since only the $\omega = 0$ value of $W$ enters into the approximated BSE. 

In practical implementations, the BSE is usually rewritten in the following matrix form in a transition space spanned by the products of occupied and unoccupied orbitals:\cite{louie2000prb,MolGWBse,bseBenchBlase}
\begin{equation}\label{BSE-matrix-form}
        \begin{bmatrix} A & B \\ -B^\dagger & -A^\dagger \\
        \end{bmatrix}
        \begin{bmatrix} X_s \\ Y_s \\
        \end{bmatrix} = 
        E_s 
        \begin{bmatrix} X_s \\ Y_s \\
        \end{bmatrix} \, .
\end{equation}
Here, $E_s$ are the optical excitation eigenvalues and ($X_s, Y_s$) are the eigenvectors. The $X_s$ and $Y_s$ are expressed in electron-hole space of the unperturbed system with elements $X_{s,ia}$ and $Y_{s,ia}$, i.e., the actual BSE wavefunctions are linear combinations of the product of (g)KS orbitals. The excitation wavefunctions $X_s, Y_s$ can be taken to be real-valued in finite (molecular) systems without an external field. Blocks $A$ and $-A^\dagger$ correspond to resonant transitions from occupied to unoccupied orbitals and the antiresonant transitions, respectively.\cite{Bechstedt16} Blocks $B$ and $-B^\dagger$ describe the coupling between blocks $A$ and $-A^\dagger$. 
In the BSE, the diagonal matrices $A(-A^\dagger)$ and off-diagonal matrices $B(-B^\dagger)$ are defined as:\cite{louie2000prb,MolGWBse,bseBenchBlase} 
\begin{equation} \label{blocka}
\begin{split}
A_{ia}^{jb} = & (\epsilon_a^{GW} - \epsilon_i^{GW})\delta_{ij}\delta_{ab}\\ & -\alpha^{S/T} (ia|V|jb) + (ij|W(\omega = 0)|ab) \, ,
\end{split} 
\end{equation} 
\begin{equation} \label{blockb}
B_{ia}^{jb}=-\alpha^{S/T} (ia|V|bj) + (ib|W(\omega = 0)|aj) \, .
\end{equation}
The indices $i$ and $j$ denote occupied states and $a$ and $b$ denote unoccupied states. $\epsilon_a^{GW}$ and $\epsilon_i^{GW}$ are the quasiparticle energies denoted as $\epsilon_l^{GW}$ in equation (\ref{gw}). The coefficient $\alpha^{S/T}$ is equal to 2 for singlet excitations and 0 for triplet excitations. 
The index conventions for the bare and screened Coulomb interactions $V$ and $W$ are as follows:\cite{MolGW}
\begin{equation}\label{Viajb}
(ia|V|jb)  = \sum_{pqrs} c^\dagger_{ip} c_{aq} c_{jr} c^\dagger_{bs} (pq|rs) \, ,
\end{equation}
\begin{equation}\label{Wijab}
(ij|W|ab)  = \sum_{pqrs} c^\dagger_{ip} c_{jq} c_{ar} c^\dagger_{bs}  (pq|W|rs) \, .
\end{equation}
$(pq|rs)$ are the two-electron integrals in a basis set representation:
\begin{equation} \label{pqrs}
(pq|rs) = \iint \frac{\varphi_p(\mathbf{r}) \varphi_q(\mathbf{r}) \varphi_r(\mathbf{r}') \varphi_s(\mathbf{r}')}{|\mathbf{r} - \mathbf{r}'|} d\mathbf{r} d\mathbf{r}' \, ,
\end{equation}
and the same convention (Mulliken notation) for $\mathbf{r}$ and $\mathbf{r}^\prime$ is used in the notation of the screened Coulomb integrals $(pq|W|rs)$ as well.
The neglect of the coupling blocks $B$($-B^\dagger$) in Eq. (\ref{BSE-matrix-form}) is known as the Tamm-Dancoff approximation (TDA).\cite{tdaPrl1998,tdaPrb1998}
In the TDA, which also we compare below, the relevant equation becomes simply
\begin{equation} \label{BSE:TDA}
A X_s = E_s X_s \, .
\end{equation}

The oscillator strength $f_s$ can be calculated from the eigenvalues and eigenvectors obtained by solving the BSE eigenvalue problem:\cite{bseOscBench}
\begin{equation} \label{osc1}
f_s = \frac{2}{3} E_s \sum_{\mu=x,y,z} (d_{s,\mu})^2 \, .
\end{equation}
$d_{s,\mu}$ can be calculated as\cite{bseOscBench}
\begin{equation} \label{osc2}
d_{s,\mu} = \sqrt{2} \sum_{ia} \langle \psi_i|\hat{\mu}|\psi_a\rangle  (X_{s,ia} + Y_{s,ia}) \, .
\end{equation}
Since we are dealing with finite systems, the dipole operator $\hat{\mu}$ is simply taken to be the position operator, i.e., $\hat{\mu}\equiv(x,y,z)$. For convenience, we reference the coordinates $x$, $y$, $z$ to the center (average of atomic positions) of the molecule.

We will also compare our observations for the BSE to analogous results for LR-TDDFT, which is widely used in chemistry. We therefore briefly recapitulate the LR-TDDFT formalism, the mathematical structure of which is similar to the BSE, albeit with a two-point kernel instead of the four-point kernel of BSE. A deeper discussion of the mathematical similarities and differences of both levels of theory is given in Ref. \citenum{reining2002review}. LR-TDDFT is often expressed as the Casida eigenvalue equation,\cite{casidaFormula} which is formally equivalent to Eq. (\ref{BSE-matrix-form}). Here, the LR-TDDFT formalism becomes
\begin{equation} \label{casida}
\Omega F_s = E_s^2 F_s \, .
\end{equation}
In LR-TDDFT, $\Omega$ is called the Casida matrix, which has the same dimension as $A$ or $B$ in Eq. (\ref{BSE-matrix-form}). $E_s^2$ are squares of the neutral many-body excitation energies and $F_s$ are the eigenvectors of this eigenvalue problem, which can also be related to the oscillator strengths via the dipole operator.\cite{casida1995} The Casida matrix can be written in a basis of products of (g)KS orbitals as
\begin{equation} \label{tddftmatrix}
 \Omega_{ia, jb}(\omega) = \delta_{i, j} \delta_{a, b} (\epsilon_a - \epsilon_i)^2 + 2\sqrt{(\epsilon_a - \epsilon_i)} K_{ia, jb}(\omega) \sqrt{(\epsilon_b - \epsilon_j)} \, ,
\end{equation}
 where $\delta$ denotes the Kronecker delta. The kernel $K_{ia, jb}$ is defined as
 \begin{equation}\label{casidaKernel}
 K_{ia, jb}(\omega) = \int\int \psi_i^\dagger(\mathbf{r}) \psi_a(\mathbf{r}) [\frac{1}{|\mathbf{r} - \mathbf{r}^\prime|} + \tilde{f}_\mathrm{xc}[n_0](\mathbf{r}, \mathbf{r}^\prime, \omega)] \psi_j(\mathbf{r}^\prime) \psi_b^\dagger(\mathbf{r}^\prime) d\textbf{r} d\textbf{r}^\prime \, .
 \end{equation}
 $\tilde{f}_\mathrm{xc}[n_0]$ is the exchange correlation kernel and $n_0$ is the (g)KS ground state electron density. The kernel is formally defined through the functional derivative of the time-dependent Kohn-Sham exchange and correlation potential $v_\mathrm{xc}[n](\textbf{r}, t)$ with respect to the time-dependent density $n(\textbf{r}^\prime, t^\prime)$ such that
\begin{equation} \label{tddft_ker}
\tilde{f}_\mathrm{xc}[n_0](\textbf{r}, \textbf{r}^\prime, \omega) = \int{ d(t-t^\prime)e^{i \omega (t-t^\prime)} \frac{\delta v_\mathrm{xc}[n](\textbf{r}, t)}{\delta n(\textbf{r}^\prime, t^\prime)} | _{n_0}} \, .
\end{equation}
In practice, the so-called adiabatic approximation\cite{FB05} is employed in Eq. (\ref{tddft_ker}), as we do here, and the exchange-correlation kernel reads
\begin{equation} \label{tddft_ker_adia}
\tilde{f}^{A}_\mathrm{xc}[n_0](\textbf{r}, \textbf{r}^\prime) =  \frac{\delta v_\mathrm{xc}[n](\textbf{r})}{\delta n(\textbf{r}^\prime)} | _{n_0} \, .
\end{equation}
This approximation makes the exchange-correlation kernel frequency-independent. 

\section{\label{sec:implementation}Implementation}
In our implementation, the two-electron Coulomb interaction in equations (\ref{blocka}), (\ref{blockb}) and (\ref{tddftmatrix}), the static screened Coulomb interaction in equations (\ref{blocka}) and (\ref{blockb}), as well as the two-electron integrals of the exchange correlation kernel in equation (\ref{tddftmatrix}), are calculated employing the RI approach.\cite{Whitten1973,Dunlap1979,Feyereisen1993,Vahtras1993,Weigend1998,RenRI2012} 
The RI represents pair products of atomic basis functions $\varphi_p(\mathbf{r}) \cdot \varphi_q(\mathbf{r})$ in terms of auxiliary basis functions (ABFs)
\begin{equation}
\varphi_p(\mathbf{r}) \varphi_q(\mathbf{r}) \approx \sum_{\mu} C^{\mu}_{pq}  P_{\mu}(\mathbf{r}) \, ,
\end{equation}
where $P_{\mu}(\mathbf{r})(\mu = 1, 2, ..., N_{aux})$ are the ABFs and $C^{\mu}_{pq}$ are the expansion coefficients. The construction of the ABFs in FHI-aims is explained in Ref. \citenum{RenRI2012} and in detail in Ref. \citenum{IhrigNJP2015}. The evaluation of the integrals (\ref{pqrs}) then reduces to
\begin{equation}
(pq|rs) \approx \sum_{\mu \nu} C^{\mu}_{pq} (\mu|\nu) C^{\nu}_{rs} \, ,\,\mathrm{with}
\end{equation}
\begin{equation}
(\mu|\nu) = \int \frac{P_{\mu}(\mathbf{r}) P_{\nu}(\mathbf{r}')}{|\mathbf{r} - \mathbf{r}'|} d\mathbf{r} d\mathbf{r}' \, .
\end{equation}
The computation of the expansion coefficients $C^{\mu}_{pq}$ requires three-center integrals involving the ABFs and the pair products of the NAOs:
\begin{equation}
C^{\mu}_{pq} = \sum_{\nu} (pq|\nu) (\nu|\mu)^{-1} \, , \, \mathrm{where}
\end{equation}
\begin{equation}
(pq|\nu) = \iint \frac{\varphi_p(\mathbf{r}) \varphi_q(\mathbf{r}) P_{\nu}(\mathbf{r}')}{|\mathbf{r} - \mathbf{r}'|} d\mathbf{r} d\mathbf{r}' \, .
\end{equation}
$(\nu|\mu)^{-1}$ denotes the inverse of the Coulomb matrix in ABF representation. Thus, the expensive computation of four-center integrals $(pq|rs)$ is reduced to the computation of much cheaper three-center and two-center ones:
\begin{equation}
(pq|rs) \approx \sum_{\mu} O^{\mu}_{pq} O^{\mu}_{rs} \, , 
\end{equation}
using 
\begin{equation}
O^{\mu}_{pq} = \sum_{\nu} C^{\nu}_{pq} (\nu|\mu)^{-1/2} \, .
\end{equation}
$(\nu|\mu)^{-1/2}$ denotes the square root of the inverse Coulomb matrix. This enables the efficient computation of the Coulomb matrix elements both in time and memory. The screened Coulomb interaction $W$ can be represented in terms of the ABFs in a similar way to the Coulomb interaction $V$:
\begin{equation}
( pq|W|rs ) = \sum_{\mu\nu} O^{\mu}_{pq} (\varepsilon^{-1})_{\mu \nu} O^{\nu}_{rs} \, .
\end{equation}
The dielectric function $\varepsilon$ can be calculated as  
\begin{equation}
\varepsilon_{\mu\nu} = \delta_{\mu\nu} - \sum_{\alpha\beta}(\mu|\alpha)^{1/2}\chi^0_{\alpha\beta} (\beta|\nu)^{1/2} \, ,
\end{equation}
where $\chi^0$ is the non-interacting density response function. In real space and for a non-spinpolarized system,  $\chi^0$ is defined as 
\begin{equation}
\chi^0(\mathbf{r},\mathbf{r}^\prime,i\omega) = 2 \sum_i \sum_a \left[ \frac{ \psi_i^\dagger(\mathbf{r}) \psi_a(\mathbf{r}) \psi_a^\dagger(\mathbf{r}^\prime) \psi_i(\mathbf{r}^\prime)}{i\omega + \epsilon_a - \epsilon_i } + c.c. \right]\, .
\end{equation}
``c.c.'' denotes the complex conjugate. We refer to Ref. \citenum{RenRI2012} for more details. The current BSE implementation in FHI-aims uses global RI.\cite{RenRI2012} In the future, use of the localized RI formalism\cite{IhrigNJP2015} is expected to facilitate scalability to larger systems as well as support for extended (periodic) systems.

\section{\label{sec:results}Results}
\subsection{\label{sec:validation}Numerical Validation}
We quantify the precision of our BSE implementation by calculating the neutral excitation energies of the molecular benchmark set published by Schreiber \textit{et al.},\cite{thiel} known in the literature as ``Thiel's set''. This set (see Fig. 1 in Ref. \citenum{thiel}) includes $N$=28 small and medium-sized organic molecules, the largest of which is naphthalene with 18 atoms. The chemical elements represented in these organic molecules are H, C, N, and O. The atomic coordinates for the included molecules are taken from the supporting information of Ref.~\citenum{thiel}. Schreiber \textit{et al.} focused on obtaining ``Best Estimate (BE)'' values for singlet and triplet excitation energies of these molecules from \textit{ab initio} calculations, including rather demanding multireference, coupled cluster or complete active space approaches of their own or from the literature. While the BE values have been used as reference data by both Thiel's and other groups for implementations, evaluation and development of a variety of methods,\cite{thielTddft,MolGWBse,bseBenchBlase} our present study is aimed at 
establishing the numerical precision of our approach at a fixed level of theory, i.e., BSE or LR-TDDFT. We therefore do not compare to the BE results, but rather compare the BSE excitation energies calculated by our present NAO-based implementation to results obtained at the same level of theory, using the MolGW code as a benchmark.  Regarding the basis set, Schreiber \textit{et al.}\cite{thiel} (and therefore, also some of the results from other methods available for comparison in the literature) employed a relatively limited basis set level for correlated calculations, the TZVP basis set by Sch\"afer \textit{et al.}\cite{tzvp92} A previous study by Bruneval \textit{et al.}\cite{MolGWBse} indicates that BSE@$G^0W^0$@DFT-B3LYP\cite{Becke1993} excitation energies for ethene and pyrrole, obtained with the TZVP basis set, overestimate the analogous results from the much larger aug-cc-pVQZ basis set\cite{Kendall1992} by about 0.45-0.65~eV. 
Therefore, the goal of our following investigation is twofold. We first establish the numerical precision of our own implementation in comparison to MolGW using the TZVP basis set.\cite{tzvp92} We then discuss basis set convergence for low-lying singlet and triplet excitations using NAO basis sets.

\begin{figure}
	\includegraphics[width=0.7\textwidth]{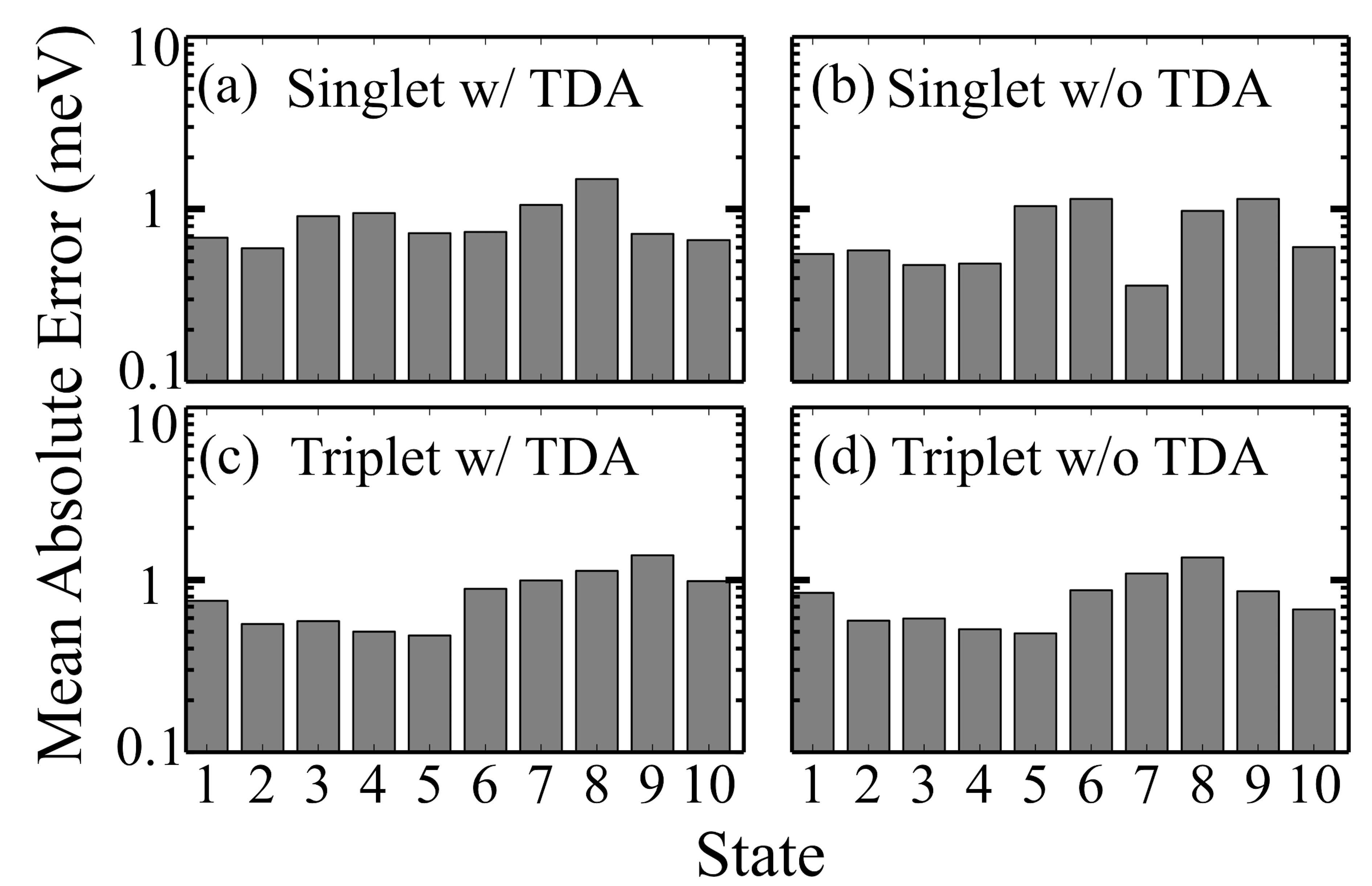}
	\caption{Mean absolute error MAE($i$) (Eq.~(\ref{mae})) of the BSE@$G^0W^0$@PBE lowest 10 singlet/triplet excitation energies from our implementation, compared with reference values from MolGW. The GTO-type TZVP basis set\cite{tzvp92} is used. The $G^0W^0$ quasiparticle energies for the FHI-aims BSE calculations are here taken from MolGW, ensuring that the comparison is specific to the BSE part of the calculations. 
	Panels (a) and (b) show the MAE of the singlet states with and without the Tamm-Dancoff approximation (TDA), respectively. Panels (c) and (d) show the MAE of the triplet states with and without TDA, respectively.}
	\label{fig:benchmark}
\end{figure}

In Figure~\ref{fig:benchmark}, we compare the numerical precision of the present BSE implementation and that in MolGW using the TZVP basis set. Specifically, we show state-resolved mean absolute error (MAE) values of the BSE-approximated energies of the lowest ten singlet and triplet excited states, respectively, of all molecules in Thiel's set. The state-resolved MAE, MAE($i$), of a given dataset $D=\{D_{i,n}\}$ in comparison to a reference set $R=\{R_{i,n}\}$ is defined as:
\begin{equation} \label{mae}
\mathrm{MAE}(i) = \frac{1}{N} \sum^N_{n = 1} |D_{i, n} - R_{i, n}| \, .
\end{equation}
$i$=1, ..., 10 is the index for the state and $n$=1, ..., $N$ is the index for the  molecules in Thiel's set. For the MAE plotted in Fig. \ref{fig:benchmark}, the dataset $D$ is the set of BSE excitation energies calculated using the FHI-aims implementation. The reference set $R$ is the set of BSE excitation energies calculated by MolGW. The PBE\cite{pbe} exchange-correlation functional is used for the initial DFT calculations and the $G^0W^0$ quasiparticle energies that enter the BSE are taken from MolGW in both sets. Panels (a) and (b) show the results for singlet excitations with and without TDA, respectively. Panels (c) and (d) show the results for triplet excitations with and without TDA, respectively. The BSE results from the present implementation agree with the MOLGW package at the level of 1 meV or below. 

\begin{figure}
	\includegraphics[width=0.5\textwidth]{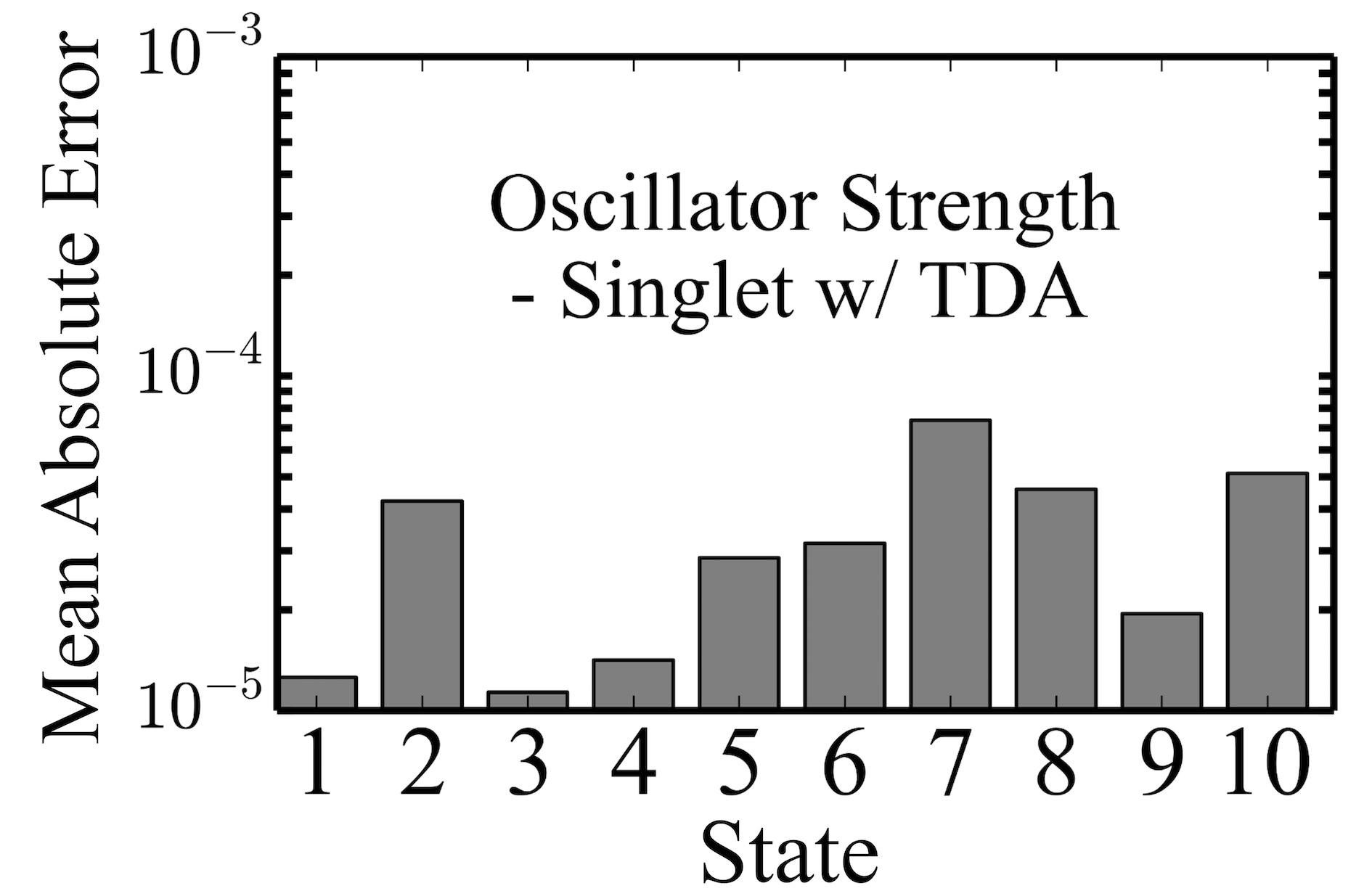}
	\caption{Mean absolute error MAE($i$) of the BSE@$G^0W^0$@PBE oscillator strengths from the present implementation compared to MolGW, using the TDA and the GTO-type TZVP basis set\cite{tzvp92} for the lowest 10 singlet excitation states, averaged over all molecules in Thiel's set. As in Figure~\ref{fig:benchmark}, the $G^0W^0$ quasiparticle energies for the FHI-aims BSE calculations are here taken from MolGW.}
	\label{fig:osc}
\end{figure}

Next, Figure~\ref{fig:osc} compares BSE oscillator strengths, Eqs. (\ref{osc1}) and (\ref{osc2}), from both implementations for singlet states in the TDA. The MAEs for the oscillator strengths are at the level of $\mathrm{10^{-4}}$ or below, i.e., numerically negligible. The actual value of the excitation energy and oscillator strength investigated in this section for all the molecules in Thiel's set can be found in the Table S1-S5 of the Supplementary Material (SM). Since the calculation results of the FHI-aims and MolGW package agrees within 1 meV for the excitation energy and $10^{-4}$ for the oscillator strength, the values in Table S1-S5 are valid for both packages within the significant digits given. In short, Figures~\ref{fig:benchmark} and \ref{fig:osc} validate our implementation.

\subsection{Effects of the frequency dependence of the self-energy in GW}
\begin{figure}
	\includegraphics[width=0.5\textwidth]{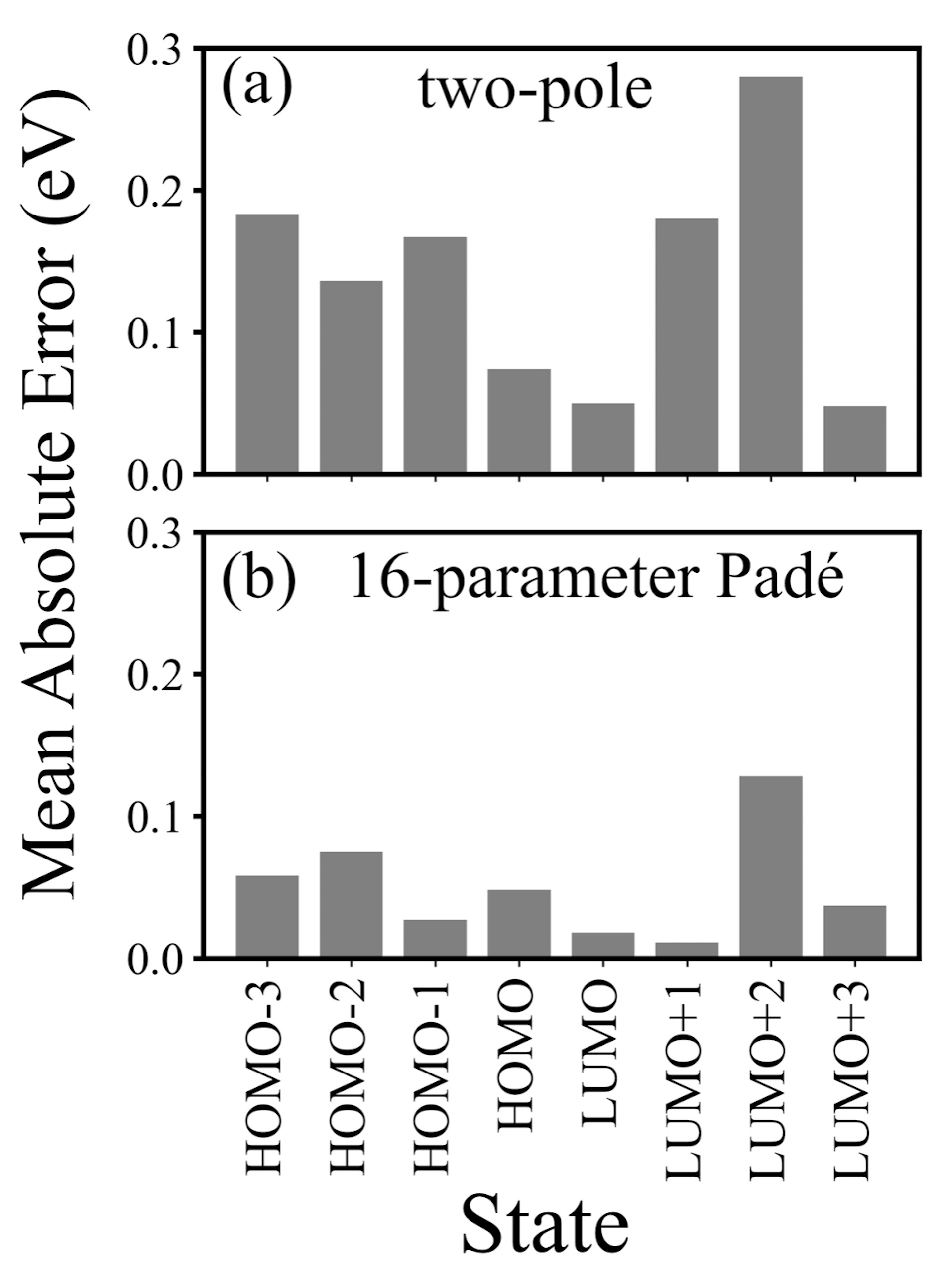}
	\caption{Impact of two different self-energy extrapolation schemes in FHI-aims on  $G^0W^0$ quasiparticle energies for the HOMO-3 to LUMO+3 states, compared to the formally exact self-energy treatment in MolGW (reference). Mean absolute errors MAE($i$) for Thiel's molecular benchmark set were computed using the GTO-type TZVP basis set\cite{tzvp92}. (a) MAE($i$) using the two-pole approximation, Eq. (\ref{2pole}). (b) MAE($i$) using the Pad\'{e} approximation, Eq. (\ref{pade}), for $N$=16.}
	\label{fig:compareGW}
\end{figure}
We next investigate how the BSE energies are impacted by different treatments of the frequency dependence of the $G^0W^0$ self-energy $\Sigma^{G^0W^0}$. In the previous section we took the $G^0W^0$ quasiparticle energies calculated by MolGW as the input for the FHI-aims BSE calculations. $G^0W^0$ calculations in MolGW employ an exact analytic treatment of $\Sigma^{G^0W^0}$ on the real axis.\cite{MolGW,gw_exact_freq} A similarly precise result is expected from the CD technique by Golze \textit{et al.}\cite{GolzeGW} In the present section, we investigate the impact of two frequently employed inexact but potentially cost-saving\cite{GolzeGW} approximations to the self-energy on the real axis, namely the two-pole approximation,\cite{godbyneeds1995} Eq. (\ref{2pole}), and the Pad\'{e} approximation,\cite{pades} Eq. (\ref{pade}), with 16 parameters.

In Figure~\ref{fig:compareGW}, we show mean absolute errors MAE($i$) of the $G^0W^0$ quasiparticle energies of the HOMO-3 to LUMO+3 states, calculated either using the two-pole approximation (Fig. \ref{fig:compareGW}(a)) or the 16-parameter Pad\'{e} approximation (Fig. \ref{fig:compareGW}(b)) and compared to the MolGW reference values. The $G^0W^0$ results are based on DFT calculations using the Perdew-Burke-Ernzerhof (PBE)\cite{pbe} exchange-correlation functional and employ the Gaussian-type TZVP basis set.\cite{tzvp92} We see that the two-pole approximation gives MAE values of up to 0.3 eV in the investigated states. Although not a small value, this must be viewed in the context of plain DFT errors (no $GW$ approximation), which are typically of the order of several eV. The 16-parameter Pad\'{e} approximation can reduce this error by a factor of two or better, as shown in Fig. \ref{fig:compareGW}(b). Most of the quasiparticle energies agree with the MolGW results within 0.1 eV, except for the state LUMO+2, where the MAE is around 0.12 eV. The performance of both approximations is closely in line with a broader analysis performed in Ref. \citenum{gw100}. However, another practical advantage of the two-pole approximation is its relative numerical simplicity and therefore its relative robustness against numerical errors, compared to the Pad\'{e} approximation. In practice, the Pad\'{e} approximation can result in serious ambiguities for individual eigenvalues $\epsilon_l^{GW}$ if there are multiple solutions for the self-consistency condition between the left and the right sides of Eq. (\ref{gw}). Thus, the two-pole approximated self-energy can be preferable for simple reasons of stability, at the price of reduced accuracy compared to a formally exact $G^0W^0$ self-energy.

\begin{figure}
	\centering
	\includegraphics[width=1.0\textwidth]{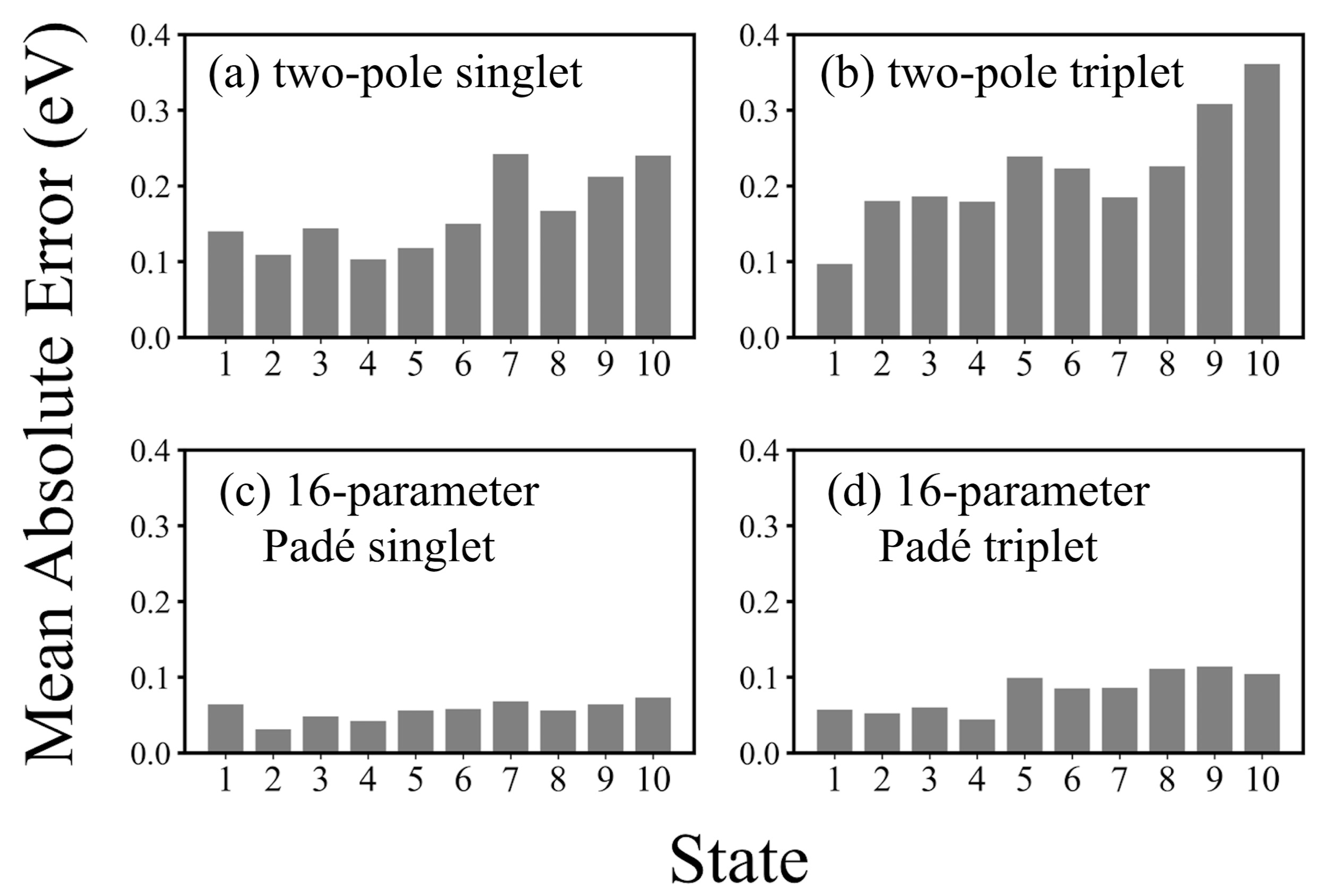}
	\caption{Mean absolute errors MAE($i$) of the lowest ten BSE@$G^0W^0$@PBE excitation energies, averaged over Thiel's set of molecules, comparing results using $G^0W^0$ eigenvalues from analytically continued self-energies and results using the formally exact $G^0W^0$ self-energy in MolGW as a reference. TZVP basis sets\cite{tzvp92} and the TDA are used in all calculations. (a) MAE($i$) of singlet states using $G^0W^0$ quasiparticle energies from the two-pole approximation, Eq. (\ref{2pole}). (b) MAE($i$) of triplet states using the two-pole approximation. (c) MAE($i$) of singlet states using $G^0W^0$ quasiparticle energies from the 16-parameter Pad\'{e} approximation, Eq. (\ref{pade}). (d) MAE($i$) of triplet states using the 16-parameter Pad\'{e} approximation.}
	\label{fig:diffgw}
\end{figure}

As shown in Figure~\ref{fig:diffgw}, the different approximate $G^0W^0$ self-energy treatments affect the BSE excitation energies, which take the $G^0W^0$ quasiparticle energies as input. We compare BSE results based on $G^0W^0$ quasiparticle energies calculated using the self-energies of Eqs. (\ref{2pole}) and (\ref{pade}) to BSE results based on the MolGW $G^0W^0$ eigenvalues with the exact analytic self-energy treatment.\cite{MolGW,gw_exact_freq} Panels (a) and (b) show the MAE($i$) values for the ten lowest singlet and triplet BSE excitation energies, respectively, using the two-pole approximation and averaged over all molecules in Thiel's set. Panels (c) and (d) show the analogous results obtained using the 16-parameter Pad\'{e} approximation. We see that the 16-parameter Pad\'{e} approximation yields smaller MAE($i$) ($\approx$0.1 eV across all investigated states) than the two-pole approximation (MAE($i$) of 0.1-0.4 eV). The difference is a direct reflection of the different MAE($i$) of the $G^0W^0$ quasiparticle energies (Figure~\ref{fig:compareGW}), which constitute the input for the BSE calculations. 

In addition to the MAE of the BSE@$G^0W^0$@PBE results between FHI-aims and MolGW, we can also define the state-dependent mean signed error MSE($i$), 
\begin{equation} \label{mse}
\mathrm{MSE}(i) = \frac{1}{N} \sum^N_{n = 1} (D_{i, n} - R_{i, n})  \, .
\end{equation}
Just as in Eq. (\ref{mae}), $i$ and $n$ are indices for state and molecules, respectively, averaging over $N$ = 28 molecules in Thiel's set. Table~\ref{relErr} summarizes the overall MSE and MAE (averaged over states $i$ in addition to averaging over molecules) of the BSE@$G^0W^0$@PBE results using the two-pole and 16-parameter Pad\'{e} approximation vs. MolGW as a reference.
As noted above, the MAE of BSE results based on quasiparticle eigenvalues from the 16-parameter Pad\'{e} approximation is less by a factor of two than the MAE of BSE results from two-pole approximated quasiparticle eigenvalues. The MSE indicates that BSE results from the two-pole quasiparticle eigenvalues overestimate the expected BSE@$G^0W^0$ excitation energies based on an exact $G^0W^0$ self-energy. 

It is interesting to compare the errors incurred from the different $G^0W^0$ self-energy treatments to the errors associated with the BSE approach itself. Bruneval \textit{et al.}\cite{MolGWBse} show that the BSE@$G^0W^0$@PBE singlet excitation energies at the TZVP basis set level give a MSE of $-$0.8 eV and a MAE of 0.8 eV compared to the BE results of Schreiber \emph{et al.}\cite{thiel} The MSE and MAE of the BSE@$G^0W^0$@PBE triplet excitation energies are around $-$1.2 eV and 1.2 eV, respectively.\cite{MolGWBse} In comparison, the error incurred through the approximate $G^0W^0$ self-energies in Table~\ref{relErr} is rather small for the two-pole approximation and negligible for the 16-parameter Pad\'{e} approximation. Additionally, the sign of the MSE from both self-energy approximations is the opposite of the MSE compared to the BE values, i.e., especially the two-pole approximation would actually reduce the MSE compared to the BE values, as a result of fortuitous error cancellation. However, this reduction should not be trusted systematically. For a true improvement over the reported small-molecule BSE excitation energies, it would obviously be preferable to pursue higher-level approaches than BSE@$G^0W^0$@PBE at the TZVP basis set level -- in terms of the DFT starting point, in terms of the theoretical treatment of the neutral excitation, and in terms of the basis set.

Even for the small-molecule systems in Thiel's set, the BSE correction still accurately captures the vast majority of the change between straight differences of HOMO and LUMO levels from $G^0W^0$ calculations and actual neutral excitation energies. This correction is much larger than the differences incurred above as a result of the approximate $G^0W^0$ self-energies. In our calculations, the BSE@$G^0W^0$@PBE results using both 16-parameter Pad\'{e} approximation and two-pole approximations reduce the lowest $G^0W^0$ HOMO-LUMO gap by about 57\%, averaged over all molecules in Thiel's set. For the He atom, essentially basis set converged results by Li \textit{et al.}\cite{hePrl2017} show a $G^0W^0$@Hartree-Fock (HF) fundamental gap of 24.69~eV, compared to significantly lower lowest-lying triplet and singlet excitation energies of 19.82~eV and 21.22~eV, respectively. Even in this extreme case of an isolated two-electron atom, these essentially basis set converged BSE@$G^0W^0$@HF results agree with the exact result to better than $\approx$0.01 Ha ($\approx$0.3 eV). For larger molecules, such as free-base porphyrin and tetraphenylporphyrin,\cite{Palummo2009} BSE@$G^0W^0$LDA excitations can match experimental absorption spectra to a similarly good degree (essentially exact within the remaining approximations made in Ref.~\cite{Palummo2009}). Here, the BSE again corrects the simple $G^0W^0$ HOMO-LUMO energy difference by several eV, much more than the magnitude of changes due to analytical corrections to the $G^0W^0$ self-energies reported above. 

\begin{table}
    \caption{Mean absolute error and mean signed error of the BSE@$G^0W^0$@PBE excitation energies using analytical continuations (two-pole or 16-parameter Pad\'{e} approximation) for the $G^0W^0$ quasiparticle eigenvalues. The values are averaged over the lowest ten states and all molecules in Thiel's set, using the exact $G^0W^0$ self-energy treatment in MolGW as a reference. Results are shown for both singlet and triplet excitation energies.}
    {\centering
    \begin{tabular}{lllll}
    \hline
    $G^0W^0$ self-energy & \multicolumn{2}{c} {Two-pole} & \multicolumn{2}{c} {16-Parameter Pad\'{e}} \\
    & Singlet & Triplet & Singlet & Triplet \\
    \hline
    MSE & 0.10 eV & 0.15 eV & 0.01 eV & 0.04 eV \\
    MAE & 0.16 eV & 0.22 eV & 0.06 eV & 0.08 eV \\
    \hline
    \end{tabular}
   }
    \label{relErr}
\end{table}

\subsection{\label{sec:basis_conv}Basis Set Convergence}
We now address the basis set convergence of the NAO basis sets for the BSE calculation, in comparison to cc-pVnZ basis sets\cite{dunning1989} and aug-cc-pVnZ basis sets\cite{Kendall1992} by Dunning and coworkers. Two types of NAO basis sets have been constructed in the context of FHI-aims. The first, denoted as ``FHI-aims-2009'', was introduced in Ref. \citenum{Blum2009} and aimed at ground-state DFT calculations. The FHI-aims-2009 basis sets come in different tiers (i.e., levels) n=1, 2, 3, 4 (in the case of H, a fourth tier does not exist and the molecular results for tier4 below employ tier3 for H). These basis sets allow one to achieve total-energy convergence corresponding to fast qualitative calculations to few-meV/atom\cite{elephant} calculations in a single, hierarchical basis set scheme (i.e., for a given element, each basis set level contains the exact basis functions from all lower-accuracy basis set levels as a subset). For first- and second-row elements, FHI-aims' ``tight'' settings employ all FHI-aims-2009 basis functions up to and including tier 2, shown in Table S6 in the SM. The second type of basis set is denoted as NAO-VCC-nZ with n = 2, 3, 4, 5.\cite{Igor2015NJP} The NAO-VCC-nZ basis functions are constructed in analogy to the cc-pVnZ correlation-consistent (cc) basis sets by Dunning,\cite{dunning1989} but employing the numerically tabulated shape of NAOs (nodeless hydrogen-like radial functions with a numerical confinement potential applied to the tails). The NAO-VCC-nZ basis sets are optimized to be suitable for converging electronic total-energy calculations based on valence-only correlation methods that include sums over unoccupied states, e.g., RPA or MP2.\cite{Igor2015NJP}
In the following, the above two types of NAO basis sets, as well as cc-pVnZ (n = 2, 3, 4, 5)\cite{dunning1989} and aug-cc-pVnZ (n = 2, 3, 4, 5) basis sets,\cite{Kendall1992} are compared to the aug-cc-pV5Z basis sets as the benchmark reference in the BSE calculations.

\begin{figure}
	\includegraphics[width=1.0\textwidth]{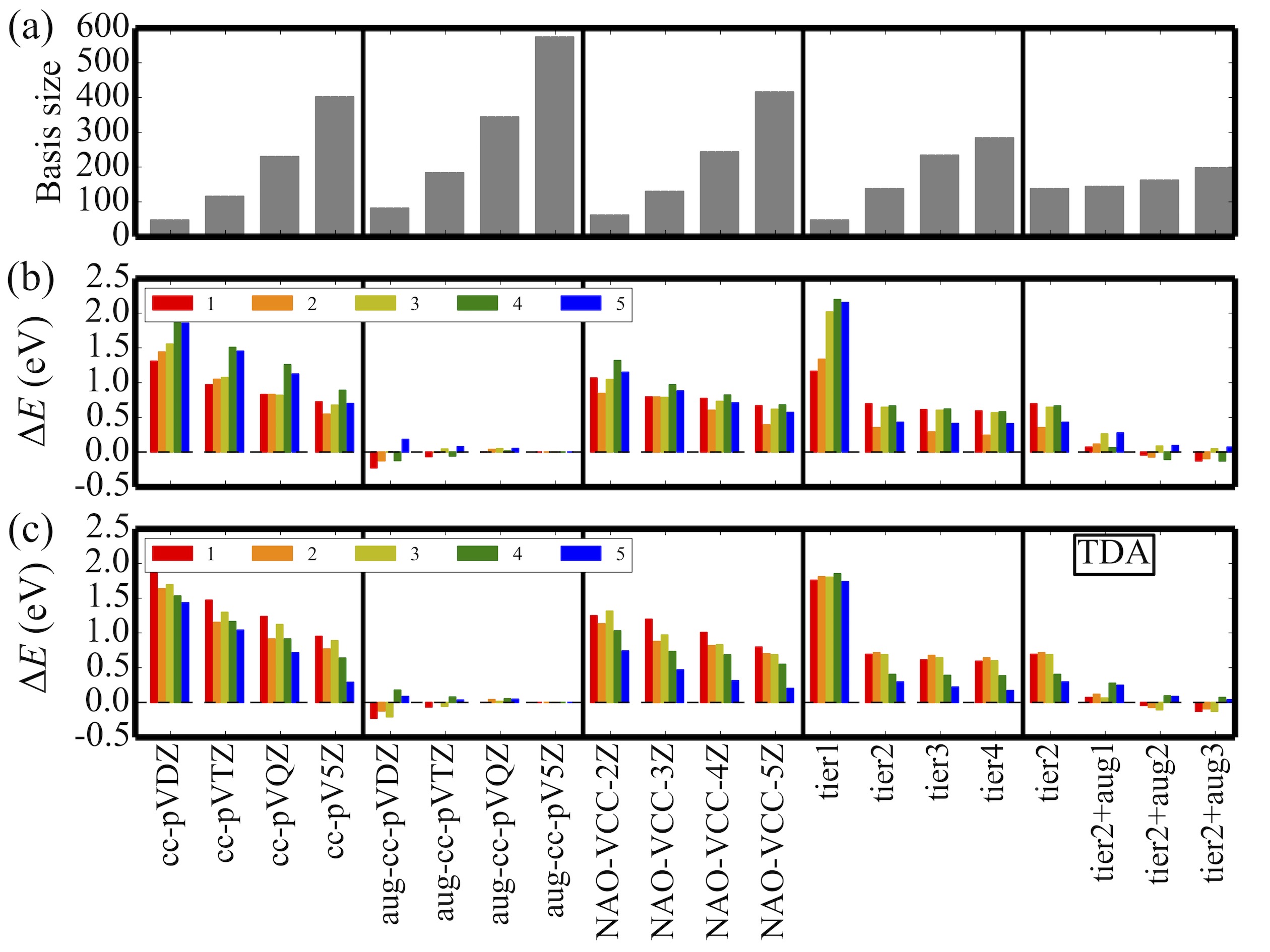}
	\caption{The difference between the BSE excitation energy computed with different basis sets and that of the aug-cc-pV5Z basis set for the lowest 5 singlet excitations of the Ethene molecule in Thiel's set. Panel (a) shows the basis size of all basis sets. Panel (b) and panel (c) give the difference of the BSE excitation energy without and with TDA, respectively. The excitation energy calculated using the aug-cc-pV5Z basis set by Dunning and coworkers\cite{Kendall1992} is used as the reference value. The ``tier'' notation corresponds to the basis set tiers of the ``FHI-aims-2009'' basis sets,\cite{Blum2009} either unmodified or with additional augmentation functions from the ``aug-cc'' basis sets as described in the text.}
	\label{fig:convergence}
\end{figure}

In Figure \ref{fig:convergence}, we show the difference between the BSE excitation energy computed with different basis sets and that of the aug-cc-pV5Z basis set for the lowest five singlet excitations of the Ethene molecule in Thiel's set. The different investigated basis set types and levels are identified on the $x$ axes of all three panels. The size of the different basis sets for the Ethene molecule is plotted on the $y$ axis in panel (a). The difference $\Delta E_i$ between the BSE excitation energy computed using different basis sets and that computed using the aug-cc-pV5Z basis set is identified on the $y$ axes in panel (b) and (c): 
\begin{equation} \label{deltaESing}
\Delta E_i = E^{\textrm{basis}}_i - E^{\textrm{aug-cc-pV5Z}}_i \, .
\end{equation}
The $\Delta E_i$ of the lowest five singlet states ($i$=1-5) are plotted in Fig. \ref{fig:convergence}(b) without the TDA and in Fig. \ref{fig:convergence}(c) with the TDA.  In both Figs. \ref{fig:convergence}(b) and (c), we see that the results obtained with the cc-pVnZ basis sets converge slowly towards the reference value as the basis size increases. The remaining discrepancy is greater than 0.5 eV even for the very expensive cc-pV5Z basis set. Although the magnitude of this discrepancy is unsatisfactory, its occurrence is not surprising. The stated reason for developing the augmented cc basis sets was an improved description of electron affinities,\cite{Kendall1992} a key constituent of the BSE@$GW$@DFT excitation energies discussed here. Accordingly, the results obtained with the aug-cc-pVnZ basis sets converge much faster. This is in line with the literature:\cite{MolGWBse, bseBenchBlase} E.g., in Fig. 3 of Ref. \citenum{MolGWBse}, the LR-TDDFT and BSE@$G^0W^0$@B3LYP\cite{Becke1993} excitation energies of the Ethene and Pyrrole molecules using the aug-cc-pVDZ and aug-cc-pVQZ basis sets show differences of less than 0.1 eV for Ethene and about 0.2 eV for Pyrrole, for both LR-TDDFT and BSE.
In Figs.~\ref{fig:convergence}(b) and (c), the results obtained with the NAO-VCC-nZ basis sets, which are constructed analogously to the cc-pVnZ basis sets, display a similarly unsatisfactory convergence behavior as that of the results from cc-pVnZ basis sets. The other type of NAO basis sets, i.e., the FHI-aims-2009 ``tier'' basis sets, behaves slightly differently compared to the NAO-VCC-nZ basis sets. The FHI-aims-2009 tier2 basis set improves the BSE excitation energies significantly compared to the FHI-aims-2009 tier1 results, in fact even slightly better than the two un-augmented cc basis set prescriptions. However, the FHI-aims-2009 tier3 and tier4 results are very similar to those obtained using the tier2 basis set, displaying no further significant improvement. 

The results discussed so far confirm the importance of the augmentation functions.
We thus extend the FHI-aims-2009 tier2 basis set with different numbers of Gaussian-type augmentation functions obtained from the aug-cc-pV5Z basis set. The label ``tier2+aug1'' in Fig. \ref{fig:convergence} denotes the basis set generated by adding the first Gaussian augmentation function from the aug-cc-pV5Z basis set (angular momentum quantum number $l$=0) to the FHI-aims-2009 tier2 basis set. Similarly, the label ``tier2+aug2'' denotes the basis set obtained by combining the FHI-aims-2009 tier2 basis set with the first two Gaussian augmentation function ($l$ = 0, 1) from the aug-cc-pV5Z basis set. We see that the addition of the augmentation functions significantly improves the accuracy of the FHI-aims-2009 tier2 results compared to the reference aug-cc-pV5Z values. Specifically, the tier2+aug2 basis set already yields essentially basis set converged values for the excitation energies of Ethene shown in Fig. \ref{fig:convergence}, with a remaining discrepancy of 0.1 eV or less compared to aug-cc-pV5Z. This conclusion is independent of whether or not the TDA is used in the BSE calculation, as shown in Fig. \ref{fig:convergence}(c). As an important main result, the tier2+aug2 basis sets can here provide rather well converged values, comparable to the aug-cc-pV5Z reference values for low-lying neutral excitations. As will be shown below, this result can be generalized to the remainder of the molecules in Thiel's set. Interestingly, the tier2+aug basis sets thus provide a recipe allowing one to use a basis set that is precise but affordable for ground-state DFT\cite{elephant} and, with a very limited modification, sufficient to achieve highly converged BSE results for low-lying excitations at the same time. For the lowest-energy excitations, which are often those of the greatest interest, we can thus use a very similar NAO basis set prescription as in ground-state DFT.

\begin{figure}
	\includegraphics[width=1.0\textwidth]{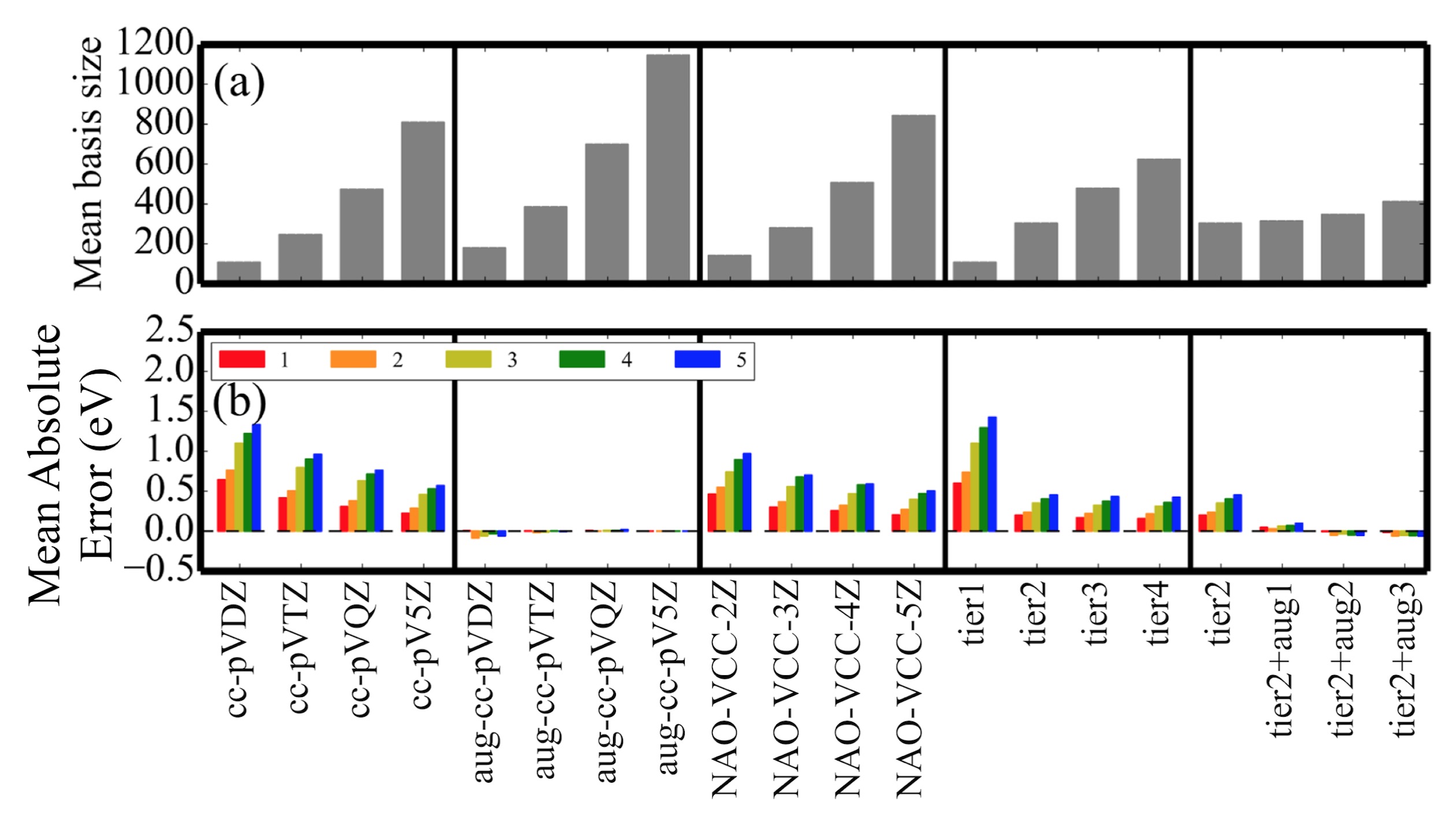}
	\caption{(a) The size of all basis sets shown, averaged over all molecules in Thiel's set (see Eq. (\ref{Eq:av-basis})). (b) MAE($i$) values of the lowest five BSE@$G^0W^0$@PBE excitation energies computed with different basis sets, using results from the aug-cc-pV5Z basis set as the reference. The TDA is employed.}
	\label{fig:all_conv}
\end{figure}

 Figure~\ref{fig:all_conv} shows the convergence of the MAE($i$) of the five lowest-lying excitation energies as a function of the size of various basis sets in BSE calculations, employing the TDA and averaged over all molecules in Thiel's set. The reference is, again, the aug-cc-pV5Z basis set. The different investigated basis set types and levels are identified on the $x$ axes of panels (a) and (b). The basis size $\bar N_{\textrm{basis}}$ of different basis sets, averaged over all molecules in Thiel's set, is plotted as the $y$ axis in Fig. \ref{fig:all_conv}(a). $\bar N_{\textrm{basis}}$ is defined as:
\begin{equation}\label{Eq:av-basis}
\bar N_{\textrm{basis}} = \frac{1}{N_{\textrm{mol}}} \sum^{N_{\textrm{mol}}}_{j = 1} N^j_{\textrm{basis}} \, .
\end{equation}
Here, $N_{\textrm{mol}}$ is the number of molecules in Thiel's set and $N^j_{\textrm{basis}}$ is the basis size for molecule $j$ in the benchmark set. 
We see that the convergence of different basis sets is similar to the earlier observations made for Ethene in Fig. \ref{fig:convergence}. Specifically, the tier2+aug2 basis set produces rather well converged results for all molecules investigated here. The lowest five singlet excitation energies calculated by BSE@$G^0W^0$@PBE using the aug-cc-pV5Z and tier2+aug2 basis sets are also listed in Table S7 in the SM for all molecules in Thiel's set.

\subsection{Convergence with respect to the BSE matrix size}\label{sec:cutoff}
In the BSE calculations presented in this work so far, we include the orbital pairs of \textit{all} occupied and unoccupied orbitals in the construction of the BSE matrix.  The dimension of the matrix problem Eq. (\ref{BSE-matrix-form}) thus grows quadratically with the basis set size and also (for a fixed basis set level) with molecular size. Solving this full BSE matrix problem produces an excitation spectrum of the studied molecule that includes very high excitation energies. In many practical applications, however, we are interested in only the low-lying part of the excitation spectrum. In such a situation, one might suspect that the high-lying unoccupied quasiparticle states do not contribute much to the low-lying optical excitations. This can, in fact, not be entirely true, since single-quasiparticle like $GW$ observables such as the ionization potential can depend significantly on high-lying parts of the spectrum being included in unoccupied-state sums in the relevant perturbation expressions.\cite{UmariPrb2011} All-electron approaches to $G^0W^0$ band gaps suffer from similar convergence issues with basis set size, specifically the basis set resolution in those regions of space that are closer to a nucleus.\cite{Jiang2016}
However, neutral excitations are not the same objects, and the question thus remains how many quasiparticle states, especially the high-lying unoccupied orbitals, should be included in the construction of the BSE matrix, in order to obtain sufficiently precise results for low-lying optical excitation energies. Previous investigations on leaving away states in the calculation of excitation energies exist. For instance, this was done in Ref.~\citenum{Palummo2009}, already referenced above. As another common example, various studies show how to efficiently select the desirable orbital space in the study of complete active space approaches.\cite{ROOS1980157, casSele2003, casSele11, casSele16} As final example, in the calculation of the electronic spectra of molecules in solution or surfaces, Besley developed an approach within LR-TDDFT and single excitation configuration interaction that limits electronic excitations to include only those between orbitals localized on the solute or adsorbant.\cite{BESLEY2004124}

\begin{table}
    \caption{The number of unoccupied states, $N_{unocc}$, averaged over all molecules in Thiel's set, when imposing threshold values $E_{cut}$=20, 40, 60, and 80 eV, or no threshold (``-'') in the BSE@$G^0W^0$@PBE calculations. The tier2+aug2 basis set is used.}
    {\centering
    \begin{tabular}{llllll}
    \hline
    Threshold (eV) & 20 & 40 & 60 & 80 & - \\
    \hline
    $N_{unocc}$ & 64 & 95 & 117 & 142 & 292 \\
    \hline
    \end{tabular}
   }\\
    \label{unoccnum}
\end{table}

\begin{figure}
	\includegraphics[width=0.5\textwidth]{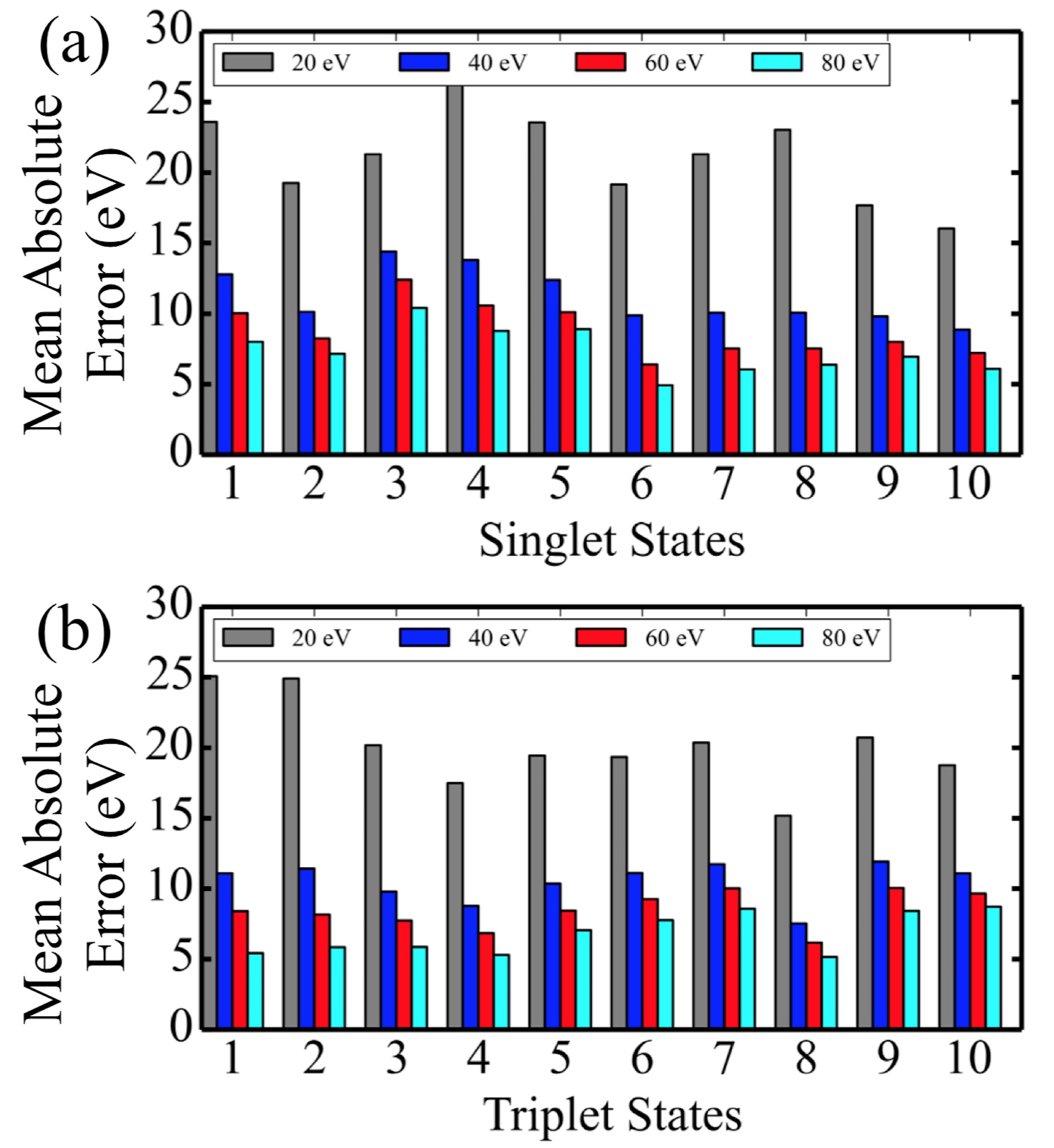}
	\caption{(a) MAE($i$) of the BSE@$G^0W^0$@PBE excitation energies of the lowest ten singlet states using different energy cutoff values $E_{cut}$ (20, 40, 60, 80 eV) for unoccupied states, using the case where all unoccupied states are included as a reference. The basis set is the tier2+aug2 basis set and the values are averaged over all molecules in Thiel's set. Panel (b) shows the analogous MAE($i$) for triplet states.}
	\label{fig:states_conv}
\end{figure}

In Figure~\ref{fig:states_conv}, we show the errors incurred in the BSE low-lying singlet and triplet excitation energies obtained when applying different values of a cutoff energy $E_{cut}$ for unoccupied states, limiting the number of states $a$ and $b$ entering the matrix construction in Eqs.~(\ref{blocka}) and (\ref{blockb}). Here, $E_{cut}$ denotes a cutoff energy above which the high-lying unoccupied quasiparticle states are omitted from the BSE matrix (however, such cutoffs are not applied in the construction of the quantities entering the BSE matrix). The average numbers of unoccupied states included for different choices of $E_{cut}$ are tabulated in Table~\ref{unoccnum}. Figure~\ref{fig:states_conv} shows MAE($i$) values for the lowest ten singlet (subfigure~\ref{fig:states_conv}a) or triplet (subfigure~\ref{fig:states_conv}b) excitation energies for the different cutoff energy values, using the tier2+aug2 basis set for all calculations and the excitation energies from the full calculation (no cutoff imposed) as a reference.
In these calculations, all occupied states are kept in the construction of the BSE matrix, i.e., no cutoff threshold is applied to the occupied quasiparticle states. Figure~\ref{fig:states_conv} shows that the error of the results calculated with $E_{cut}$=20 eV is about 20-30 meV. Setting $E_{cut}$=40 eV yields excitation energies with an error closer to 10 meV. Larger $E_{cut}$ values of 60 and 80 eV lead to further small improvements. In view of the remaining errors of these calculations (due to level of theory for underlying DFT, quasiparticle energies, neutral excitation formalism itself), these results suggest that one may use 40 eV as a threshold beyond which the impact of high-lying unoccupied quasiparticle states becomes negligible in BSE calculations of low-lying excitations. This can reduce the computational effort significantly because of the reduction of number of states needed in construction of the BSE matrix. Specifically, the time and memory complexity of constructing the BSE matrix in Eqs. (\ref{blocka}) and (\ref{blockb}) scale as ~${N}_{occ}^2 \times {N}_{unocc}^2$, where $N_{occ}$ and $N_{unocc}$ denote the number of occupied and unoccupied (g)KS single-particle states, respectively. By setting $E_{cut}$=40 eV and for the tier2+aug2 basis sets, the number of unoccupied states is reduced to about 1/3 of the analogous number if no threshold energy values are used (see Table \ref{unoccnum}). Additionally, the formal effort for solving the full BSE, Eq. (\ref{BSE-matrix-form}), would scale as $O(N^6)$, where $N$ is a measure of system size, due to the same considerations of how $N_{occ}$ and $N_{unocc}$ impact the matrix dimension. While imposing $E_{cut}$ will not reduce the formal scaling, the actual computational effort will nevertheless be reduced substantially in the limit of large systems where the BSE solution must eventually dominate. In short, both the time and the memory requirements of the BSE calculation of low-lying excitations are expected to be reduced significantly by imposing $E_{cut}$, without sacrificing much accuracy.

\subsection{Comparison to LR-TDDFT}\label{comparewithTddft}
In this section, we first investigate the basis set convergence of LR-TDDFT excitation energies for the molecules in Thiel's set, using the strategy already employed for the BSE in Section \ref{sec:basis_conv}. The LR-TDDFT in FHI-aims was implemented following Eqs. (\ref{casida}-\ref{tddft_ker_adia}). The exchange-correlation kernel used is the LDA, employing the parametrization of the correlation energy by Perdew and Wang,\cite{ks1965,pwlda} provided by the Libxc library.\cite{libxc1, libxc2}
Note that the LR-TDDFT formalism leaves one with the freedom to choose different prescriptions of the XC functional for (i) the self-consistent solutions of single-particle orbitals and energies and (ii) the XC kernel $f_{xc}$ (Eq. (\ref{tddft_ker_adia})) used in the actual LR-TDDFT construction. In this section, the exchange correlation functional for the self-consistent solutions of single-particle orbitals is PBE.\cite{pbe} We will use the notation ``LR-TDDFT-LDA@PBE'' in the following to denote this approach.

\begin{figure}
	\includegraphics[width=1.0\textwidth]{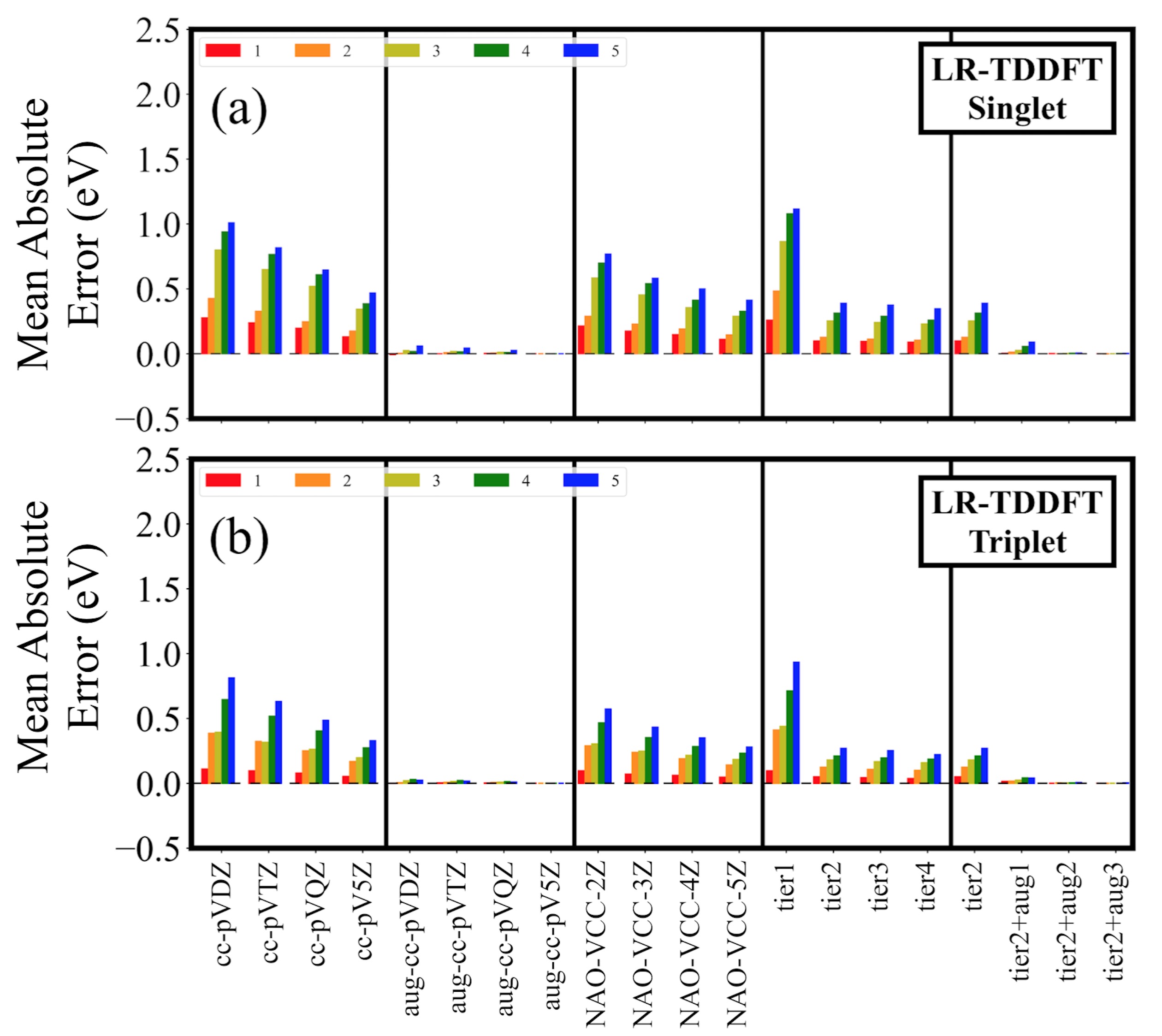}
	\caption{(a) MAE($i$) of LR-TDDFT-LDA@PBE excitation energies for the lowest five singlet excitation states, averaged for all molecules in Thiel's set, computed with different basis sets and using the aug-cc-pV5Z basis set as a reference. Panel (b) shows the analogous MAE($i$) for triplet states.}
	\label{fig:tddft_bench}
\end{figure}

Fig. \ref{fig:tddft_bench} provides state-dependent MAE($i$) values for the lowest five LR-TDDFT excitation energies computed with different basis sets, averaged over Thiel's set. Excitation energies computed with the aug-cc-pV5Z basis sets are used as reference values. The different investigated basis set types and levels are identified on the $x$ axes of both panels (a) and (b). 
Fig.~\ref{fig:tddft_bench}(a) for singlet states and Fig. \ref{fig:tddft_bench}(b) for triplet states show essentially identical behavior. The overall convergence pattern associated with all the basis sets investigated here is also very similar to the behavior seen for the BSE in Figure~\ref{fig:all_conv}. Excitation energy values derived from both the cc-pVnZ basis sets and the NAO-VCC-nZ basis sets converge slowly towards the reference value as the basis size increases. The largest error is about 0.5 eV for the very expensive cc-pV5Z basis set for singlet excitation energies and 0.3 eV for triplet excitation energies. The aug-cc-pVnZ basis sets, again, converge much faster. As for the BSE, the FHI-aims-2009 basis sets improve significantly towards the reference results from tier1 to tier2, but not further using tier3 and tier4. Finally, by adding augmentation functions to the FHI-aims-2009 tier2 basis sets, one can obtain well converged LR-TDDFT results by including only one or two augmentation basis functions. The tier2+aug2 basis set, already discussed for BSE calculations above, leads to compellingly good basis set convergence as a production recipe. The lowest five singlet and triplet excitation energies calculated by LR-TDDFT-LDA@PBE using aug-cc-pV5Z and tier2+aug2 basis set are provided in Tables S8 and S9 in the SM for all molecules in Thiel's set.

\begin{figure}
	\includegraphics[width=1.0\textwidth]{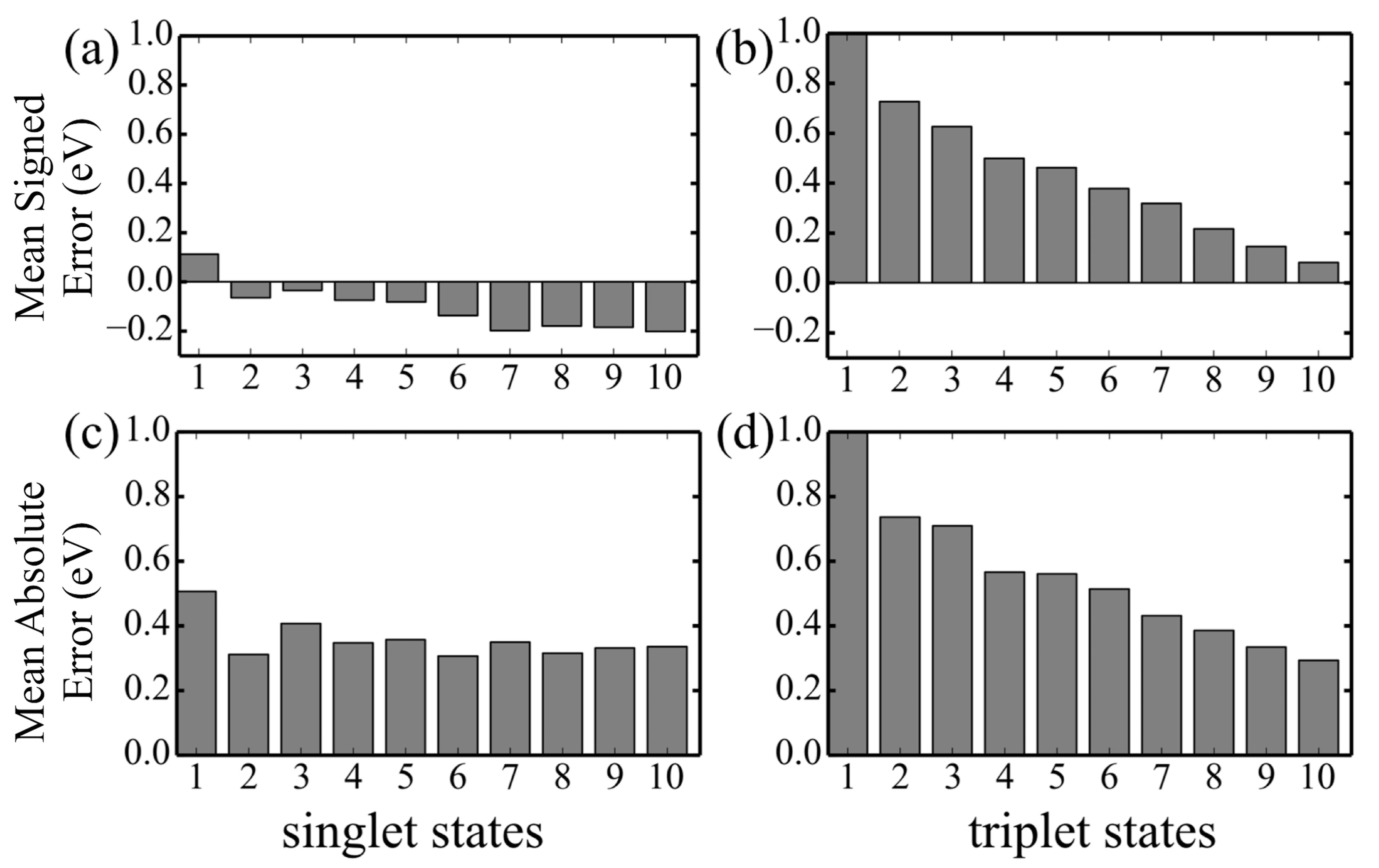}
	\caption{Mean signed errors MSE($i$) and mean absolute errors MAE($i$) between LR-TDDFT-LDA@PBE and BSE@$G^0W^0$@PBE (reference) results, averaged for all molecules in Thiel's set, using the tier2+aug2 basis set in the TDA for the lowest ten singlet and triplet states. Panels (a) and (b) show the MSE of the singlet and triplet states, respectively. Panels (c) and (d) show the MAE of the singlet and triplet states, respectively.}
	\label{fig:tdlda_bse}
\end{figure}

We finally compare the results of BSE@$G^0W^0$@PBE with those of LR-TDDFT-LDA@PBE, implemented in FHI-aims and using the tier2+aug2 basis set validated above. While the primary focus of the present work is numerical and basis set convergence, we provide this comparison here because LR-TDDFT is widely used in quantum chemistry as a computationally efficient method for optical excitation calculations. We note that similar comparisons can be found in the literature,\cite{MolGWBse, bseBenchBlase} albeit not using the same basis sets. In our comparison, the underlying DFT calculations for both BSE and LR-TDDFT are carried out using the PBE\cite{pbe} exchange-correlation functional. To maintain consistency, the TDA is employed in the BSE calculations, as this is widely done for LR-TDDFT calculations as well. Figure~\ref{fig:tdlda_bse} shows the MSE($i$) and the MAE($i$) between LR-TDDFT-LDA@PBE results and the BSE@$G^0W^0$@PBE results, averaged over all molecules in Thiel's set. The BSE@$G^0W^0$@PBE results are taken as the reference. The lowest ten singlet and triplet excitation energies are compared. The MSE between LR-TDDFT-LDA@PBE and BSE@$G^0W^0$@PBE results for singlet and triplet excitations states is plotted in Fig. \ref{fig:tdlda_bse} (a) and (b), respectively. The MAE between LR-TDDFT-LDA@PBE and BSE@$G^0W^0$@PBE results for singlet and triplet excitations states is plotted in Fig. \ref{fig:tdlda_bse} (c) and (d), respectively. 
We see from Fig. \ref{fig:tdlda_bse} that the MAE between LR-TDDFT-LDA@PBE and BSE@$G^0W^0$@PBE for singlet excitation energies is less than 0.5 eV, where as the MAE for triplet excitations energies can be as large as 1.0 eV. Singlet excitation energy values obtained from LR-TDDFT-LDA@PBE tend to be lower than those obtained from BSE@$G^0W^0$@PBE, whereas LR-TDDFT-LDA@PBE triplet excitation energies appear to be larger by up to 1~eV than those obtained from BSE@$G^0W^0$@PBE. Previous studies of BSE and LR-TDDFT show that the excitation energies computed by BSE and LR-TDDFT depend highly on the DFT starting point.\cite{MolGWBse,bseBenchBlase} Bruneval \textit{et al.}\cite{MolGWBse} compare the MSE of both BSE$G^0W^0$@B3LYP and LR-TDDFT@B3LYP excitation energies with the BEs of Thiel's set, showing that the MSE of LR-TDDFT is about 0.4 eV lower than the MSE of BSE.\cite{MolGWBse, thielTddft} However, there are several differences between their comparison and the comparison shown in this work. First, the dataset used by Bruneval and coworkers are the BEs of Thiel's set,\cite{thiel} which contains 103 singlet and 63 triplet excitation energies. In our work, we have a larger dataset to include the lowest ten singlet and triplet states of each molecule in Thiel's set. Second, the BSE and LR-TDDFT calculations analyzed in this section rely on the TDA, which is not employed in the comparison performed by Bruneval et al.\cite{MolGWBse} Finally, we here use a basis set that is essentially converged for both BSE and LR-TDDFT calculations. Another set of comparable results are therefore those of Jacquemin and coworkers,\cite{bseBenchBlase} who compare the BSE@$G^0W^0$@PBE0 and LR-TDDFT@PBE0 in a benchmark paper using the aug-cc-pVTZ basis, which has similar convergence behavior as the tier2+aug2 basis set used here. Different MAE values between BSE@$G^0W^0$@PBE0 and LR-TDDFT@PBE0 excitation energies are reported for different categories of molecules in Thiel's set: 0.27 eV for unsaturated aliphatic hydrocarbons; 0.51 eV for aromatic compounds; 0.37 eV for Aldehydes, ketones, and amides; 0.47 eV for nucleobases. The reported values are comparable to the values that we find in Fig. \ref{fig:tdlda_bse}, in the range of 0.2 - 1.0 eV for different states of singlet and triplet excitation energies.

\begin{figure}
	\includegraphics[width=1.0\textwidth]{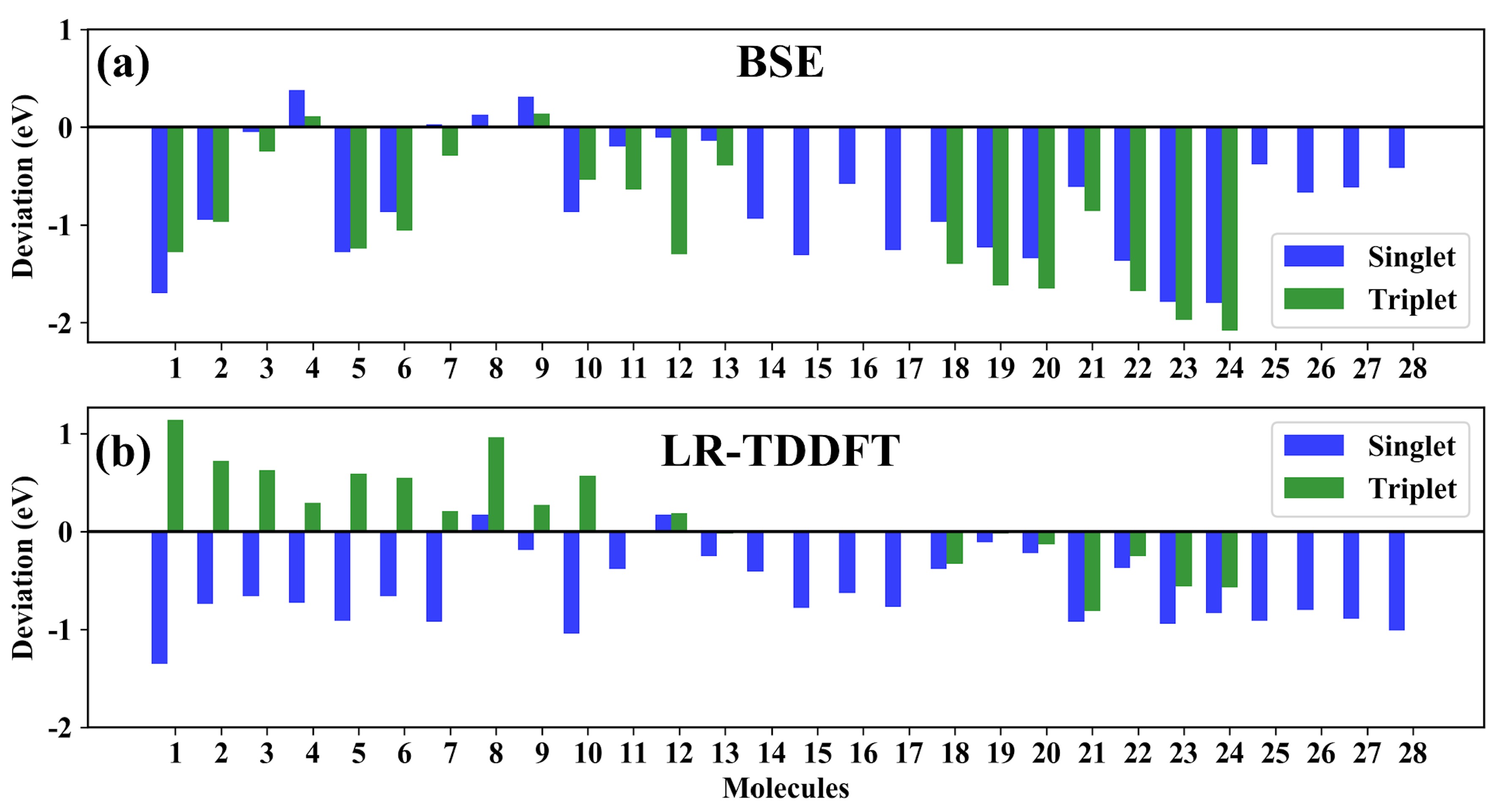}
	\caption{Deviations of the BSE@$G^0W^0$@PBE and LR-TDDFT-LDA@PBE results for the lowest singlet and triplet excitation energies of each molecule in Thiel's set, compared with the BE values.\cite{thiel} Panels (a) and (b) show the comparison for BSE and LR-TDDFT, respectively. The molecule indices follow the order in the original paper by Schreiber \textit{et al.}\cite{thiel}. The BE values for the lowest triplet state of molecules 14-18 and 25-28 are not available, therefore the comparison of the lowest triplet state is omitted for these molecules.}
	\label{fig:compare_bte}
\end{figure}

\subsection{Remarks on Time and Memory Cost of the BSE and LR-TDDFT Implementation}
While the present work does not include a performance optimization of either the BSE or the LR-TDDFT implementations reported above, we provide some individual timings indicative of the computational cost of our (not fully optimized) BSE and LR-TDDFT implementations. The results should only be understood as qualitative indicators of the current implementation, since no dedicated optimization was carried out, but nevertheless give some idea of the relative cost of different steps at present and indicate avenues for future work in our own implementation. In the present section, we consider a series of acene molecules of increasing size: benzene, naphathalene and anthracene. Both BSE and LR-TDDFT calculations were performed using the ``tier2+aug'' basis sets described in Sections \ref{sec:basis_conv} and \ref{comparewithTddft}, respectively, and employing the TDA. 

In Table \ref{tab:timings40}, we show the timings of BSE@$G^0W^0$@PBE and LR-TDDFT-LDA@PBE performed for the three acene molecules selected. We apply a cutoff energy $E_{cut}$ = 40 eV to limit the number of unoccupied states entering the BSE and Casida matrix construction as described in Section \ref{sec:cutoff}. The calculations are performed using a single node with 44 cores (Intel Xeon, Broadwell microarchitecture, 2.4 GHz) on the Dogwood cluster at University of North Carolina, Chapel Hill. 
The total time is decomposed into three parts, \textit{i.e.}, RI basis setup denoted by ``Basis'' (precomputed three-center and two-center integrals in Sec. \ref{sec:implementation}), BSE/LR-TDDFT matrix construction or building denoted by ``Build Mat.'' and solving the BSE/LR-TDDFT matrix in the TDA as eigenvalue problems in Eqs. \ref{BSE:TDA} and \ref{casida}, denoted by ``Solver''. The ``Build Mat.'' timing accounts for the computational effort in building the BSE matrix as outlined in Equations \ref{blocka} and \ref{blockb}, versus the LR-TDDFT matrix in Equation \ref{tddftmatrix}. We note that the timings comprise the BSE/LR-TDDFT step only, whereas the timings for the preceding steps are not counted, such as the $G^0W^0$ and DFT-PBE steps in the BSE@$G^0W^0$@PBE calculation and the DFT-PBE step in the LR-TDDFT-LDA@PBE calculation. Table \ref{tab:timings40} shows that the majority of the total timing is attributed to the product basis setup step for both BSE and LR-TDDFT calculations. Both the total timing and the timings for each step are comparable for BSE and LR-TDDFT calculations, which is expected due to the similar formalisms. However, as will be seen below in Table \ref{tab:timingsFull}, the difference of timings for the matrix building step between BSE and TDDFT becomes more significant in our present implementations if the matrix size is increased.
\begin{table}
    \centering
	\caption{Timings of BSE@$G^0W^0$@PBE and LR-TDDFT-LDA@PBE calculations performed for the neutral singlet excitations of benzene, naphathalene and anthracene. The ``tier2+aug'' basis set is used and a cutoff energy $E_{cut}$ = 40 eV for unoccupied states is applied. Calculations are performed using one node with 44 cores on the Dogwood cluster at UNC-Chapel Hill (Intel Broadwell microarchitecture, 2.4 GHz). All values are given in seconds (s).}
    \begin{tabular}{ c c c c c c c c c }
    \hline
     & \multicolumn{4}{c} {BSE@$G^0W^0$@PBE} & \multicolumn{4}{c} {LR-TDDFT-LDA@PBE} \\
    \hline
	Molecules & Basis & Build Mat. & Solver & Total & Basis & Build Mat. & Solver & Total \\
	\hline
	Benzene & 270 & 9 & 1 & 280 & 293 & 7 & 1 & 301 \\
	Naphathalene & 1200 & 44 & 5 & 1249 & 1184 & 39 & 5 & 1228 \\
	Anthracene & 3089 & 142 & 16 & 3247 & 2907 & 129 & 19 & 3055 \\
	\hline
    \end{tabular}\\
    \label{tab:timings40}
\end{table}

To demonstrate how the cutoff energy $E_{cut}$ = 40 eV reduces the computational timings, we show in Table \ref{tab:timingsFull} the timings for the same systems without using any cutoff energy for both BSE@$G^0W^0$@PBE and LR-TDDFT-LDA@PBE calculations. All other computational details remain the same as used to obtain Table \ref{tab:timings40}. We see in Table \ref{tab:timingsFull} that by using all the unoccupied states in the BSE and LR-TDDFT calculations, the timings for building and solving matrix steps increase significantly compared to Table \ref{tab:timings40}. In contrast, the timings for the product basis setup step stays almost unchanged, which is expected since the number of basis functions and auxiliary basis functions is not affected by applying the cutoff energy to limit the number of unoccupied states. As mentioned above, the matrix building step in the present implementation reveals a difference between the BSE and the LR-TDDFT timings in Table \ref{tab:timingsFull}. Specifically, the matrix building step for LR-TDDFT requires the calculations of the kernel $K_{ia, jb}$ (Equation \ref{tddftmatrix} and \ref{casidaKernel}). In our parallel implementation, the state indices $i, a, j, b$ are distributed among different processors, as are the RI two- and three-center integrals, and the calculation of $K_{ia, jb}$ requires significant inter-processor communication to get the correct state indices. For BSE, the analogous inter-processor communication has to be conducted twice in the calculations of $(ia|V|jb)$ and $(ij|W|ab)$, because the state index order is different in $(ia|V|jb)$ and $(ij|W|ab)$ and thus they are calculated in separate steps. As a result the timing of the matrix building step in the BSE is significantly larger than that in the LR-TDDFT calculations. 

Finally, we verify how the cutoff energy $E_{cut}$ = 40 eV reduces the memory requirements for the BSE and LR-TDDFT matrices. In Table \ref{tab:memory}, we compare the memory used to store the BSE or LR-TDDFT matrix using all unoccupied states and the memory requirements for the matrices when applying $E_{cut}$ = 40 eV, indicating a reduction by a factor of $\sim$9-11. This is consistent with Section \ref{sec:cutoff}, where $E_{cut}$ = 40 eV was shown to reduce the number of unoccupied states to about 1/3 of the full number of states if no cutoff energy values are used. Since the BSE and LR-TDDFT matrix size scales as ~${N}_{occ}^2 \times {N}_{unocc}^2$ (Eqs. (\ref{blocka}) and (\ref{blockb}), the overall reduction amounts to $(1/3)^2$=$1/9$, which is confirmed by the memory reduction shown in Table \ref{tab:memory}.

\begin{table}
    \centering
	\caption{Timings as in Table \ref{tab:timings40}, but without any energy cutoff applied for unoccupied states.}
    \begin{tabular}{ c c c c c c c c c }
    \hline
     & \multicolumn{4}{c} {BSE@$G^0W^0$@PBE} & \multicolumn{4}{c} {LR-TDDFT-LDA@PBE} \\
    \hline
	Molecules & Basis & Build Mat. & Solver & Total & Basis & Build Mat. & Solver & Total \\
	\hline
	Benzene & 276 & 31 & 7 & 313 & 292 & 22 & 7 & 321 \\
	Naphathalene & 1205 & 314 & 72 & 1592 & 1200 & 173 & 71 & 1443 \\
	Anthracene & 3175 & 1531 & 407 & 5113 & 3010 & 824 & 405 & 4239 \\
	\hline
    \end{tabular}\\
    \label{tab:timingsFull}
\end{table}

\begin{table}
    \centering
	\caption{Memory (in Megabytes) used to store the BSE or LR-TDDFT matrix in the BSE@$G^0W^0$@PBE or LR-TDDFT-LDA@PBE calculations of neutral singlet excitations of benzene, naphathalene and anthracene. The column labeled by ``Full'' denotes the memory required if all unoccupied states enter into the BSE or LR-TDDFT matrix. The column labeled by ``$E_{cut}$ = 40 eV'' denotes the memory required when applying the cutoff energy of 40 eV for unoccupied states. The column labeled ``Ratio'' denotes the ratio of the value in the ``Full'' column over the value in the ``$E_{cut}$ = 40 eV'' column. Other computational details are the same as those in Table \ref{tab:timings40} and \ref{tab:timingsFull}.}
    \begin{tabular}{ c c c c c c c }
    \hline
     & \multicolumn{3}{c} {BSE@$G^0W^0$@PBE} & \multicolumn{3}{c} {LR-TDDFT-LDA@PBE} \\
    \hline
	Molecules & Full & $E_{cut} = 40 $eV & Ratio & Full & $E_{cut} = 40 $eV & Ratio \\
	\hline
	Benzene & 400 & 45 & 8.9 & 400 & 49 & 8.2 \\
	Naphathalene & 2549 & 261 & 9.8 & 2549 & 282 & 9.0 \\
	Anthracene & 9129 & 876 & 10.4 & 9129 & 979 & 9.3 \\
	\hline
    \end{tabular}\\
    \label{tab:memory}
\end{table}

\section{\label{sec:conclusion}Conclusions}
We describe an implementation of the Bethe-Salpeter Equation approach to neutral excitations in small molecules based on numeric atom-centered basis sets in an all-electron electronic structure framework (the FHI-aims code). Benchmarks performed using Thiel's set of small molecules\cite{thiel} demonstrate the numerical correctness of the implementation. Mean absolute errors of less than 1~meV are obtained compared to the reference values computed using the MolGW code when exactly the same basis set and underlying technical approximations are used. We next investigate the impact of analytical approximations to the $G^0W^0$ self-energy (the two-pole and 16-parameter Pad\'{e} approximations), which impact the $G^0W^0$ quasiparticle energies entering the BSE. The MAE of the BSE@$G^0W^0$@PBE results with these analytical approximations is around 0.05 - 0.20 eV, compared with the exact $G^0W^0$ self-energy used in the  MolGW reference code. The 16-parameter Pad\'{e} approximation is more precise than the two-pole approximation where it can be used, but the two-pole approximation offers an overall numerically more stable avenue. Ultimately, the differences due to either approximation are smaller than typical basis set errors and errors due to the level of theory itself as assessed in other benchmark publications. The basis set convergence behavior of the predicted low-lying excitations is investigated for the cc-pVnZ,\cite{dunning1989} FHI-aims-2009,\cite{Blum2009} NAO-VCC-nZ,\cite{Igor2015NJP} and aug-cc-pVnZ\cite{Kendall1992} literature basis sets, as well as for a simple modification of the FHI-aims-2009 tier2 basis set by adding two augmentation functions from the aug-cc basis sets,\cite{Kendall1992} called ``tier2+aug2''. For both BSE@$G^0W^0$@PBE and LR-TDDFT-LDA@PBE, adequate precision requires the use of augmentation functions, as expected from the literature. The ``tier2+aug2'' basis set provides high precision for both BSE and LR-TDDFT calculations while remaining applicable in production calculations. Furthermore, the convergence is investigated with respect to the number of unoccupied states included in the BSE or LR-TDDFT matrix construction. A threshold of $E_{cut}$=40 eV is suggested, above which the unoccupied states are discarded in the construction of either the BSE or the LR-TDDFT matrix. This threshold significantly reduces the time and memory consumption while maintaining high precision of the results for low-lying excitations. Finally, BSE@$G^0W^0$@PBE and LR-TDDFT-LDA@PBE results are compared using the tier2+aug2 basis set for Thiel's set of molecules. The difference between BSE@$G^0W^0$@PBE and LR-TDDFT-LDA@PBE is quantified by a MAE in the range of 0.2 - 1.0 eV for different singlet and triplet states calculated for molecules in Thiel's set. In agreement with the literature, deviations from higher-level ``best estimate'' values are of a similar magnitude; one likely mitigation strategy is the selection of a better starting-point density functional for BSE@$G^0W^0$@DFT.

\section*{Supplementary Material}
See supplementary material for: Excitation energies of the molecules in Thiel's set using the BSE with and without the TDA; corresponding oscillator strengths of singlets within the TDA; definition of the ``tier2'' basis sets and numerical settings; basis set convergence of excitation energies of the molecules in Thiel's set using the BSE and LR-TDDFT.

\begin{acknowledgments}
This work was financially supported by the NSF under Awards No. DMR-1729297 and No. DMR-1728921, as well as through the Research Triangle MRSEC (DMR-11-21107). An award of computer time was provided by the Innovative and Novel Computational Impact on Theory and Experiment (INCITE) program. This research used resources of the Argonne Leadership Computing Facility (ALCF), which is a DOE Office of Science User Facility supported under Contract DE-AC02-06CH11357. We would like to thank the University of North Carolina at Chapel Hill and the Research Computing group for providing computational resources and support that have contributed to these research results. XR acknowledges the financial support from Chinese National Science Foundation (Grant No. 11574283, 11874335) and the Max Planck Partner Group project.
\end{acknowledgments}


\providecommand{\noopsort}[1]{}\providecommand{\singleletter}[1]{#1}%

\end{document}


\renewcommand{\arraystretch}{0.7}
\renewcommand{\thetable}{S\arabic{table}}

\title[Supplementary Material: BSE]
  {Supplementary Material: Ab Initio Bethe-Salpeter Equation Approach to Neutral Excitations in Molecules with Numeric Atom-Centered Orbitals}
  
\author{Chi Liu}
\affiliation
{Department of Chemistry, Duke University, Durham, North Carolina 27708, United States}
\author{Jan Kloppenburg}
\affiliation{Institute of Condensed Matter and Nanoscience, Universit\'e Catholique de Louvain, Louvain-la-Neuve, 1348, Belgium}
\author{Xinguo Ren}
\affiliation{CAS Key Laboratory of Quantum Information, University of Science and Technology of China, Hefei, Anhui 230026, China}
\author{Heiko Appel}
\affiliation{Max Planck Institute for the Structure and Dynamics of Matter, Center for Free Electron Laser Science, 22761 Hamburg, Germany}
\author{Yosuke Kanai}
\affiliation
{Department of Chemistry, University of North Carolina, Chapel Hill, North Carolina 27599, United States}
\author{Volker Blum}
\email
{volker.blum@duke.edu}
\affiliation
{Department of Mechanical Engineering and Materials Science, Duke University, Durham, North Carolina 27708, United States}
\altaffiliation
{Department of Chemistry, Duke University, Durham, North Carolina 27708, United States}

\date{\today}
             
\maketitle

\section{Excitation Energy and Oscillator Strength of Thiel's Set}

\begin{table}[h]
    \centering
	\caption{Neutral excitation energies of the lowest 10 singlet states for all molecules in Thiel's molecular benchmark set calculated by BSE@$G^0W^0$@PBE with the Tamm-Dancoff approximation (TDA) and using the TZVP basis set.\cite{tzvp92} The exchange-correlation functional used is PBE\cite{pbe}. The atomic coordinates for all molecules are taken from the supporting information of Ref. \citenum{thiel}. The $GW$ quasiparticle energies for the FHI-aims BSE calculations shown here are taken from the MolGW code.\cite{MolGW,gw_exact_freq} The difference between the BSE results calculated by these two packages is on the order of 1 meV.}
    \begin{tabular}{ c c c c c c c c c c c }
	\hline
    Molecules & 1 & 2 & 3 & 4 & 5 & 6 & 7 & 8 & 9 & 10 \\
	\hline
	Ethene & 7.70 & 7.88 & 8.26 & 8.52 & 8.63 & 9.14 & 10.23 & 10.25 & 10.54 & 11.24 \\
	E-Butadiene & 6.12 & 6.48 & 6.83 & 6.84 & 6.97 & 7.21 & 7.80 & 7.81 & 8.27 & 8.63 \\
	all-E-Hexatriene & 4.47 & 4.94 & 6.08 & 6.13 & 6.30 & 6.35 & 6.43 & 6.47 & 6.66 & 6.84 \\
	all-E-Octatetraene & 3.60 & 4.20 & 5.37 & 5.45 & 5.47 & 5.65 & 5.96 & 6.08 & 6.19 & 6.26 \\
	Cyclopropene & 5.98 & 6.59 & 6.96 & 7.66 & 7.71 & 7.96 & 8.71 & 9.01 & 9.02 & 9.21 \\
	Cyclopentadiene & 5.05 & 5.94 & 6.23 & 7.12 & 7.14 & 7.17 & 7.32 & 7.55 & 7.68 & 7.85 \\
	Norbornadiene & 4.91 & 5.54 & 6.50 & 6.81 & 7.04 & 7.05 & 7.12 & 7.42 & 7.52 & 7.55 \\
	Benzene & 4.59 & 5.88 & 6.83 & 6.83 & 7.07 & 7.07 & 7.27 & 7.27 & 7.31 & 7.31 \\
	Naphthalene & 3.69 & 4.01 & 4.82 & 5.18 & 5.66 & 5.73 & 5.78 & 5.87 & 6.13 & 6.21 \\
	Furan & 5.85 & 6.00 & 6.61 & 7.35 & 7.38 & 7.73 & 7.86 & 8.01 & 8.08 & 8.11 \\
	Pyrrole & 5.36 & 5.73 & 6.06 & 6.18 & 6.76 & 6.88 & 7.34 & 7.42 & 7.62 & 7.75 \\
	Imidazole & 5.54 & 5.75 & 5.98 & 6.58 & 6.61 & 6.83 & 6.85 & 7.18 & 7.45 & 7.84 \\
	Pyridine & 4.04 & 4.21 & 4.64 & 5.94 & 6.79 & 6.92 & 6.98 & 7.18 & 7.20 & 7.29 \\
	Pyrazine & 3.25 & 3.83 & 4.56 & 4.69 & 5.18 & 6.36 & 7.14 & 7.24 & 7.56 & 7.74 \\
	Pyrimidine & 3.45 & 3.73 & 4.83 & 4.83 & 5.03 & 6.25 & 7.10 & 7.11 & 7.45 & 7.61 \\
	Pyridazine & 2.79 & 3.26 & 4.54 & 4.65 & 5.04 & 5.87 & 6.00 & 6.38 & 6.89 & 7.04 \\
	s-Triazine & 3.54 & 3.70 & 3.70 & 3.78 & 5.23 & 6.81 & 7.14 & 7.14 & 7.54 & 7.54 \\
	s-Tetrazine & 1.45 & 2.60 & 3.31 & 4.30 & 4.51 & 4.65 & 5.17 & 5.39 & 6.38 & 6.80 \\
	Formaldehyde & 2.99 & 7.28 & 7.97 & 8.85 & 8.91 & 10.08 & 10.56 & 10.65 & 11.26 & 12.17 \\
	Acetone & 3.26 & 6.30 & 7.95 & 8.05 & 8.16 & 8.25 & 8.31 & 8.86 & 8.94 & 9.28 \\
	p-Benzoquinone & 3.44 & 4.09 & 4.24 & 4.85 & 5.53 & 5.56 & 5.63 & 5.99 & 6.04 & 6.63 \\
	Formamide & 4.45 & 6.27 & 6.57 & 7.23 & 7.45 & 7.91 & 8.34 & 8.62 & 8.97 & 9.49 \\
	Acetamide & 4.48 & 6.04 & 6.30 & 7.01 & 7.56 & 7.61 & 8.05 & 8.37 & 8.73 & 9.00 \\
	Propanamide & 4.50 & 6.08 & 6.30 & 6.89 & 7.49 & 7.52 & 7.72 & 8.06 & 8.63 & 8.78 \\
	Cytosine & 3.93 & 4.19 & 4.80 & 5.36 & 5.49 & 5.84 & 5.86 & 6.16 & 6.20 & 6.51 \\
	Thymine & 3.69 & 4.52 & 4.88 & 5.30 & 5.32 & 5.67 & 5.78 & 5.89 & 6.59 & 6.75 \\
	Uracil & 4.12 & 4.34 & 4.65 & 4.73 & 4.98 & 5.33 & 5.58 & 5.63 & 5.73 & 5.79 \\
    Adenine & 4.09 & 4.34 & 4.65 & 4.68 & 4.94 & 5.35 & 5.51 & 5.57 & 5.66 & 5.72 \\
	\hline
    \end{tabular}\\
    \label{tab:singtda}
\end{table}

\newpage
\begin{table}[h]
    \centering
	\caption{Neutral excitation energies of the lowest 10 singlet states for all molecules in Thiel's molecular benchmark set calculated by BSE@$G^0W^0$@PBE without TDA. Other computational details are the same as described in Table S1.}
    \begin{tabular}{ c c c c c c c c c c c }
	\hline
    Molecules & 1 & 2 & 3 & 4 & 5 & 6 & 7 & 8 & 9 & 10 \\
	\hline
    Ethene & 7.25 & 7.69 & 7.85 & 8.50 & 8.61 & 9.10 & 10.21 & 10.22 & 10.53 & 11.19 \\
	E-Butadiene & 5.28 & 6.26 & 6.82 & 6.83 & 6.95 & 7.19 & 7.79 & 7.79 & 8.22 & 8.25 \\
	all-E-Hexatriene & 4.17 & 4.46 & 5.39 & 6.02 & 6.07 & 6.34 & 6.38 & 6.42 & 6.47 & 6.64 \\
	all-E-Octatetraene & 3.47 & 3.58 & 4.47 & 5.13 & 5.40 & 5.65 & 5.95 & 6.07 & 6.13 & 6.20 \\
	Cyclopropene & 5.95 & 6.09 & 6.95 & 7.62 & 7.66 & 7.94 & 8.65 & 8.93 & 8.98 & 9.16 \\
	Cyclopentadiene & 4.47 & 5.82 & 6.22 & 7.11 & 7.13 & 7.16 & 7.29 & 7.51 & 7.67 & 7.78 \\
	Norbornadiene & 4.59 & 5.47 & 6.49 & 6.52 & 6.60 & 7.00 & 7.04 & 7.42 & 7.47 & 7.52 \\
	Benzene & 4.56 & 5.54 & 6.31 & 6.31 & 6.82 & 6.82 & 7.25 & 7.26 & 7.30 & 7.30 \\
	Naphthalene & 3.67 & 3.73 & 4.81 & 5.14 & 5.18 & 5.39 & 5.43 & 5.59 & 6.13 & 6.20 \\
	Furan & 5.48 & 5.79 & 6.61 & 7.31 & 7.33 & 7.37 & 7.71 & 7.85 & 7.85 & 7.97 \\
	Pyrrole & 5.35 & 5.66 & 5.73 & 6.04 & 6.75 & 6.87 & 7.12 & 7.32 & 7.40 & 7.43 \\
	Imidazole & 5.51 & 5.72 & 5.75 & 6.22 & 6.53 & 6.78 & 6.83 & 7.16 & 7.36 & 7.43 \\
	Pyridine & 3.95 & 4.20 & 4.57 & 5.61 & 6.34 & 6.42 & 6.74 & 7.08 & 7.17 & 7.20 \\
	Pyrazine & 3.16 & 3.82 & 4.46 & 4.47 & 5.17 & 5.92 & 7.13 & 7.15 & 7.21 & 7.22 \\
	Pyrimidine & 3.40 & 3.71 & 4.75 & 4.76 & 4.99 & 5.87 & 6.51 & 6.72 & 7.13 & 7.59 \\
	Pyridazine & 3.22 & 3.22 & 4.48 & 4.58 & 5.01 & 5.52 & 5.93 & 6.24 & 6.42 & 6.46 \\
	s-Triazine & 3.53 & 3.66 & 3.66 & 3.72 & 5.20 & 6.40 & 6.93 & 6.93 & 7.07 & 7.07 \\
	s-Tetrazine & 3.16 & 3.16 & 3.16 & 4.23 & 4.45 & 4.46 & 5.08 & 5.36 & 5.93 & 6.42 \\
	Formaldehyde & 2.93 & 7.23 & 7.87 & 8.36 & 8.88 & 9.52 & 10.53 & 10.64 & 11.26 & 12.14 \\
	Acetone & 3.22 & 6.29 & 7.80 & 7.98 & 8.14 & 8.24 & 8.31 & 8.55 & 8.84 & 9.17 \\
	p-Benzoquinone & 3.25 & 4.06 & 4.21 & 4.25 & 5.32 & 5.55 & 5.60 & 5.86 & 5.94 & 6.32 \\
	Formamide & 4.42 & 6.17 & 6.54 & 7.19 & 7.31 & 7.44 & 8.31 & 8.61 & 8.90 & 9.45 \\
	Acetamide & 4.46 & 6.01 & 6.26 & 6.69 & 7.50 & 7.55 & 7.99 & 8.37 & 8.71 & 8.99 \\
	Propanamide & 4.47 & 6.04 & 6.26 & 6.65 & 7.32 & 7.49 & 7.70 & 8.06 & 8.63 & 8.76 \\
	Cytosine & 3.72 & 3.72 & 4.16 & 4.58 & 5.35 & 5.49 & 5.63 & 5.75 & 5.84 & 6.19 \\
	Thymine & 3.67 & 3.67 & 4.26 & 4.87 & 5.11 & 5.29 & 5.66 & 5.66 & 5.77 & 6.33 \\
	Uracil & 4.10 & 4.23 & 4.43 & 4.71 & 4.94 & 5.23 & 5.35 & 5.57 & 5.62 & 5.70 \\
    Adenine & 4.17 & 4.25 & 4.44 & 4.65 & 4.97 & 5.25 & 5.36 & 5.55 & 5.63 & 5.66 \\
	\hline
    \end{tabular}\\
    \label{tab:singnotda}
\end{table}

\newpage
\begin{table}[h]
    \centering
	\caption{Neutral excitation energies of the lowest 10 triplet states for all molecules in Thiel's molecular benchmark set calculated by BSE@$G^0W^0$@PBE with TDA. Other computational details are the same as described in Table S1.}
    \begin{tabular}{ c c c c c c c c c c c }
	\hline
    Molecules & 1 & 2 & 3 & 4 & 5 & 6 & 7 & 8 & 9 & 10 \\
	\hline
	Ethene & 3.77 & 7.25 & 7.37 & 8.16 & 8.18 & 8.38 & 9.29 & 9.65 & 9.86 & 9.86 \\
	E-Butadiene & 2.56 & 4.20 & 6.45 & 6.63 & 6.73 & 6.88 & 6.97 & 7.29 & 7.36 & 7.88 \\
	all-E-Hexatriene & 1.82 & 2.66 & 3.99 & 4.32 & 5.52 & 5.93 & 5.98 & 6.23 & 6.25 & 6.28 \\
	all-E-Octatetraene & 1.40 & 2.03 & 3.49 & 3.54 & 4.17 & 5.05 & 5.30 & 5.56 & 5.60 & 5.71 \\
	Cyclopropene & 3.64 & 5.45 & 6.41 & 7.05 & 7.08 & 7.42 & 7.59 & 7.91 & 8.26 & 8.50 \\
	Cyclopentadiene & 2.46 & 4.06 & 6.07 & 6.08 & 6.72 & 6.83 & 6.93 & 6.95 & 6.97 & 7.42 \\
	Norbornadiene & 2.90 & 3.34 & 5.35 & 6.02 & 6.34 & 6.50 & 6.79 & 6.91 & 7.24 & 7.28 \\
	Benzene & 3.53 & 3.94 & 3.94 & 4.25 & 6.06 & 6.06 & 6.67 & 6.67 & 6.87 & 6.97 \\
	Naphthalene & 2.37 & 3.22 & 3.45 & 3.46 & 3.78 & 4.23 & 4.77 & 4.94 & 5.02 & 5.32 \\
	Furan & 3.26 & 4.39 & 5.09 & 6.08 & 6.41 & 6.81 & 6.87 & 7.12 & 7.38 & 7.44 \\
	Pyrrole & 3.57 & 4.45 & 4.92 & 5.26 & 5.52 & 5.75 & 6.48 & 6.60 & 7.08 & 7.12 \\
	Imidazole & 3.76 & 4.64 & 4.90 & 5.05 & 5.66 & 5.92 & 6.00 & 6.55 & 6.63 & 6.86 \\
	Pyridine & 3.19 & 3.54 & 3.68 & 4.00 & 4.06 & 4.64 & 6.10 & 6.18 & 6.31 & 6.80 \\
	Pyrazine & 2.46 & 3.25 & 3.61 & 3.69 & 3.71 & 4.65 & 5.03 & 5.92 & 6.54 & 6.82 \\
	Pyrimidine & 2.89 & 3.40 & 3.73 & 3.78 & 4.18 & 4.26 & 4.64 & 4.92 & 6.37 & 6.73 \\
	Pyridazine & 1.86 & 2.80 & 3.38 & 3.61 & 3.93 & 4.05 & 4.56 & 4.70 & 5.58 & 5.69 \\
	s-Triazine & 3.09 & 3.24 & 3.24 & 3.41 & 4.15 & 4.58 & 4.58 & 4.92 & 6.36 & 6.36 \\
	s-Tetrazine & 0.63 & 2.21 & 2.24 & 3.17 & 3.45 & 3.65 & 3.90 & 4.26 & 4.67 & 5.02 \\
	Formaldehyde & 2.21 & 4.60 & 6.44 & 6.61 & 8.12 & 8.29 & 9.71 & 10.31 & 11.19 & 11.55 \\
	Acetone & 2.59 & 4.98 & 6.11 & 6.96 & 7.78 & 7.94 & 7.94 & 8.18 & 8.45 & 8.79 \\
	p-Benzoquinone & 1.92 & 2.30 & 3.48 & 3.54 & 3.68 & 4.08 & 5.07 & 5.10 & 5.23 & 5.24 \\
	Formamide & 3.85 & 4.19 & 6.02 & 6.04 & 6.67 & 7.20 & 7.87 & 7.93 & 8.39 & 8.80 \\
	Acetamide & 3.91 & 4.28 & 5.75 & 5.87 & 6.87 & 7.17 & 7.24 & 7.64 & 8.01 & 8.20 \\
	Propanamide & 3.93 & 4.30 & 5.75 & 5.92 & 6.91 & 7.06 & 7.19 & 7.47 & 7.90 & 7.96 \\
	Cytosine & 2.64 & 3.24 & 3.76 & 3.77 & 4.90 & 5.19 & 5.19 & 5.32 & 5.40 & 6.11 \\
	Thymine & 2.76 & 3.24 & 3.80 & 4.28 & 4.49 & 4.98 & 5.17 & 5.46 & 5.58 & 5.93 \\
	Uracil & 2.86 & 3.63 & 3.78 & 3.81 & 4.17 & 4.26 & 4.40 & 4.50 & 4.73 & 4.93 \\
    Adenine & 2.81 & 3.60 & 3.70 & 3.82 & 4.12 & 4.15 & 4.29 & 4.47 & 4.67 & 4.93 \\
	\hline
    \end{tabular}\\
    \label{tab:triptda}
\end{table}

\newpage
\begin{table}[h]
    \centering
	\caption{Neutral excitation energies of the lowest 10 triplet states for all molecules in Thiel's molecular benchmark set calculated by BSE@$G^0W^0$@PBE without TDA. Other computational details are the same as described in Table S1.}
    \begin{tabular}{ c c c c c c c c c c c }
	\hline
    Molecules & 1 & 2 & 3 & 4 & 5 & 6 & 7 & 8 & 9 & 10 \\
	\hline
	Ethene & 3.60 & 7.23 & 7.36 & 8.14 & 8.16 & 8.35 & 9.25 & 9.63 & 9.77 & 9.80 \\
	E-Butadiene & 2.37 & 4.11 & 6.44 & 6.63 & 6.72 & 6.87 & 6.95 & 7.27 & 7.34 & 7.86 \\
	all-E-Hexatriene & 1.56 & 2.53 & 3.91 & 4.30 & 5.48 & 5.92 & 5.97 & 6.21 & 6.25 & 6.25 \\
	all-E-Octatetraene & 1.11 & 1.90 & 3.47 & 3.48 & 4.10 & 4.97 & 5.29 & 5.55 & 5.59 & 5.70 \\
	Cyclopropene & 3.49 & 5.44 & 6.38 & 7.02 & 7.06 & 7.33 & 7.58 & 7.88 & 8.22 & 8.49 \\
	Cyclopentadiene & 2.30 & 3.97 & 6.04 & 6.08 & 6.70 & 6.81 & 6.91 & 6.94 & 6.96 & 7.41 \\
	Norbornadiene & 2.77 & 3.18 & 5.33 & 6.00 & 6.34 & 6.48 & 6.78 & 6.89 & 7.23 & 7.27 \\
	Benzene & 3.30 & 3.91 & 3.91 & 4.21 & 6.00 & 6.01 & 6.67 & 6.67 & 6.85 & 6.96 \\
	Naphthalene & 3.19 & 3.19 & 3.33 & 3.43 & 3.69 & 4.19 & 4.76 & 4.90 & 5.00 & 5.28 \\
	Furan & 3.09 & 4.32 & 5.01 & 6.01 & 6.40 & 6.79 & 6.83 & 7.11 & 7.36 & 7.43 \\
	Pyrrole & 3.42 & 4.40 & 4.82 & 5.26 & 5.45 & 5.73 & 6.47 & 6.59 & 7.07 & 7.11 \\
	Imidazole & 3.63 & 4.58 & 4.86 & 4.96 & 5.65 & 5.90 & 5.93 & 6.54 & 6.62 & 6.85 \\
	Pyridine & 3.15 & 3.28 & 3.65 & 3.96 & 4.04 & 4.61 & 6.03 & 6.09 & 6.30 & 6.79 \\
	Pyrazine & 2.43 & 3.20 & 3.52 & 3.57 & 3.68 & 4.57 & 5.02 & 5.89 & 6.48 & 6.80 \\
	Pyrimidine & 2.86 & 3.38 & 3.49 & 3.74 & 4.16 & 4.22 & 4.62 & 4.89 & 6.32 & 6.67 \\
	Pyridazine & 1.81 & 2.77 & 3.12 & 3.57 & 3.90 & 4.00 & 4.52 & 4.69 & 5.57 & 5.61 \\
	s-Triazine & 3.05 & 3.22 & 3.22 & 3.39 & 3.93 & 4.55 & 4.55 & 4.88 & 6.33 & 6.33 \\
	s-Tetrazine & 0.55 & 2.18 & 2.19 & 3.12 & 3.17 & 3.62 & 3.87 & 4.20 & 4.66 & 5.00 \\
	Formaldehyde & 2.17 & 4.41 & 6.41 & 6.56 & 8.08 & 8.27 & 9.68 & 10.30 & 11.19 & 11.52 \\
	Acetone & 2.55 & 4.83 & 6.10 & 6.92 & 7.77 & 7.93 & 7.93 & 8.17 & 8.43 & 8.76 \\
	p-Benzoquinone & 1.71 & 2.21 & 3.36 & 3.52 & 3.66 & 3.98 & 5.05 & 5.07 & 5.20 & 5.23 \\
	Formamide & 3.81 & 4.07 & 6.00 & 6.02 & 6.65 & 7.19 & 7.83 & 7.92 & 8.38 & 8.78 \\
	Acetamide & 3.87 & 4.16 & 5.73 & 5.86 & 6.76 & 7.16 & 7.23 & 7.63 & 7.97 & 8.19 \\
	Propanamide & 3.90 & 4.19 & 5.73 & 5.91 & 6.88 & 6.98 & 7.17 & 7.46 & 7.89 & 7.92 \\
	Cytosine & 3.17 & 3.67 & 3.73 & 4.86 & 5.14 & 5.18 & 5.32 & 5.38 & 6.11 & 6.16 \\
	Thymine & 3.21 & 3.69 & 4.22 & 4.47 & 4.93 & 5.17 & 5.45 & 5.57 & 5.88 & 6.46 \\
	Uracil & 2.73 & 3.57 & 3.69 & 3.79 & 4.11 & 4.24 & 4.33 & 4.47 & 4.69 & 4.91 \\
    Adenine & 2.58 & 3.51 & 3.52 & 3.72 & 4.02 & 4.10 & 4.19 & 4.30 & 4.60 & 4.86 \\
	\hline
    \end{tabular}\\
    \label{tab:tripnotda}
\end{table}

\newpage
\begin{table}[h]
    \centering
	\caption{Oscillator strength of the lowest 10 singlet states for all molecules in Thiel's molecular benchmark set calculated by BSE@$G^0W^0$@PBE with TDA. Other computational details are the same as described in Table S1.}
    \begin{tabular}{ c c c c c c c c c c c }
	\hline
    Molecules & 1 & 2 & 3 & 4 & 5 & 6 & 7 & 8 & 9 & 10 \\
	\hline
	Ethene & 0.032 & 0.000 & 0.524 & 0.000 & 0.000 & 0.000 & 0.000 & 0.000 & 0.000 & 0.308 \\
	E-Butadiene & 1.031 & 0.000 & 0.001 & 0.000 & 0.000 & 0.029 & 0.000 & 0.000 & 0.000 & 0.000 \\
	all-E-Hexatriene & 0.000 & 1.670 & 0.000 & 0.010 & 0.000 & 0.000 & 0.000 & 0.008 & 0.000 & 0.207 \\
	all-E-Octatetraene & 0.000 & 2.451 & 0.000 & 0.139 & 0.000 & 0.000 & 0.000 & 0.000 & 0.000 & 0.000 \\
	Cyclopropene & 0.001 & 0.071 & 0.000 & 0.000 & 0.025 & 0.003 & 0.069 & 0.127 & 0.119 & 0.316 \\
	Cyclopentadiene & 0.133 & 0.018 & 0.000 & 0.000 & 0.000 & 0.023 & 0.000 & 0.003 & 0.000 & 0.002 \\
	Norbornadiene & 0.000 & 0.192 & 0.000 & 0.000 & 0.019 & 0.000 & 0.000 & 0.002 & 0.192 & 0.511 \\
	Benzene & 0.000 & 0.002 & 0.001 & 0.182 & 0.000 & 0.018 & 0.000 & 0.011 & 0.001 & 0.340 \\
	Naphthalene & 0.003 & 0.000 & 0.066 & 0.123 & 0.011 & 0.023 & 0.001 & 0.005 & 0.003 & 0.344 \\
	Furan & 0.005 & 0.000 & 0.017 & 0.005 & 0.003 & 0.492 & 0.158 & 0.000 & 0.000 & 0.545 \\
	Pyrrole & 0.006 & 0.000 & 0.000 & 0.092 & 0.000 & 0.110 & 0.000 & 0.000 & 0.069 & 0.276 \\
	Imidazole & 0.005 & 0.000 & 0.000 & 0.019 & 0.007 & 0.031 & 0.053 & 0.443 & 0.504 & 0.001 \\
	Pyridine & 0.005 & 0.000 & 0.000 & 0.008 & 0.006 & 0.012 & 0.005 & 0.002 & 0.457 & 0.116 \\
	Pyrazine & 0.000 & 0.000 & 0.000 & 0.017 & 0.000 & 0.000 & 0.000 & 0.000 & 0.431 & 0.430 \\
	Pyrimidine & 0.005 & 0.000 & 0.000 & 0.000 & 0.000 & 0.040 & 0.000 & 0.012 & 0.001 & 0.000 \\
	Pyridazine & 0.000 & 0.000 & 0.000 & 0.000 & 0.876 & 0.876 & 0.013 & 0.000 & 0.000 & 0.000 \\
	s-Triazine & 0.000 & 0.012 & 0.005 & 0.000 & 0.016 & 0.014 & 0.173 & 0.000 & 0.027 & 0.011 \\
	s-Tetrazine & 0.000 & 0.058 & 0.000 & 0.000 & 0.000 & 0.167 & 1.726 & 0.000 & 0.000 & 0.000 \\
	Formaldehyde & 0.000 & 0.098 & 0.003 & 0.003 & 0.038 & 0.399 & 0.000 & 0.011 & 0.000 & 0.054 \\
	Acetone & 0.000 & 0.024 & 0.007 & 0.000 & 0.014 & 0.000 & 0.000 & 0.100 & 0.027 & 0.024 \\
	p-Benzoquinone & 0.000 & 0.000 & 0.000 & 0.372 & 0.000 & 0.000 & 0.000 & 0.008 & 0.000 & 0.190 \\
	Formamide & 0.001 & 0.001 & 0.008 & 0.124 & 0.001 & 0.506 & 0.002 & 0.000 & 0.006 & 0.054 \\
	Acetamide & 0.001 & 0.016 & 0.006 & 0.138 & 0.000 & 0.114 & 0.104 & 0.000 & 0.018 & 0.005 \\
	Propanamide & 0.000 & 0.014 & 0.007 & 0.097 & 0.000 & 0.165 & 0.052 & 0.000 & 0.005 & 0.069 \\
	Cytosine & 0.031 & 0.002 & 0.077 & 0.001 & 0.000 & 0.114 & 0.005 & 0.630 & 0.001 & 0.001 \\
	Thymine & 0.000 & 0.093 & 0.000 & 0.000 & 0.082 & 0.000 & 0.000 & 0.152 & 0.212 & 0.327 \\
	Uracil & 0.000 & 0.043 & 0.187 & 0.001 & 0.001 & 0.005 & 0.179 & 0.014 & 0.000 & 0.008 \\
    Adenine & 0.000 & 0.043 & 0.193 & 0.001 & 0.001 & 0.005 & 0.138 & 0.014 & 0.001 & 0.006 \\
	\hline
    \end{tabular}\\
    \label{tab:oscsingtda}
\end{table}

\newpage
\section{Definition of the FHI-aims-2009 ``Tier2'' Basis Set and Numerical Settings Used}

\begin{table}[h]
    \caption{Basis functions (radial functions) in the FHI-aims-2009 ``Tier2'' basis set and other computational settings used for the elements H, C, N, and O.}
    {\centering
    \begin{tabular}{lllll}
        \hline
        & H & C & N & O \\
        \hline
        minimal & 1s & [He]+2s2p & [He]+2s2p & [He]+2s2p  \\
        \hline
        \textit{tier} 1 & H(2s,2.1) & H(2p,1.7) & H(2p,1.8) & H(2p,1.8) \\
        & H(2p,3.5) & H(3d,6.0) & H(3d,6.8) & H(3d,7.6) \\
        & & H(2s,4.9) & H(3s,5.8) & H(3s, 6.4) \\
        \hline
        \textit{tier} 2 & H(1s,0.85) & H(4f,9.8) & H(4f,10.8) & H(4f,11.6) \\
        & H(2p,3.7) & H(3p,5.2) & H(3p,5.8) & H(3p,6.2) \\
        & H(2s,1.2) & H(3s,4.3) & H(1s,0.8) & H(3d,5.6) \\
        & H(3d,7.0) & H(5g,14.4) & H(5g,16) & H(5g,17.6) \\
        & & H(3d,6.2) & H(3d,4.9) & H(1s, 0.75) \\
        \hline
        Outermost radial function radius (\AA) & 6 & 6 & 6 & 6 \\
        l\_hartree & 8 & 8 & 8 & 8 \\
        radial\_base & 24 & 34 & 35 & 36 \\
        radial\_multiplier & 6 & 6 & 6 & 6 \\
        \hline
    \end{tabular}
   }\\
    The first line (``minimal'') summarizes the free-atom radial functions of occupied states (noble-gas configuration of the core and quantum numbers of the additional valence radial functions), all of which are part of the basis set. ``H($nl$, $z$)'' denotes a hydrogen-like basis function for the bare Coulomb potential $z/r$, including its radial and angular momentum quantum numbers, $n$ and $l$. This nomenclature follows the convention employed in Table 1 of reference\citenum{Blum2009}. The maximum radius of each radial function is also listed as the row of ``Outermost radial function radius'' for each species. The parameter ``l\_hartree'' is the expansion order of the Hartree potential around each species. The parameter ``radial\_base'', together with the parameter ``radial\_multiplier'', defines the number of radial integration shells used for three-dimensional integrals, as is shown in more detail in the appendix of Ref. \citenum{Igor2013NJP}.
    \label{tight}
\end{table}

\newpage
\section{Basis set convergence of the BSE in FHI-aims}

\begin{table}[h]
    \centering
	\caption{Neutral excitation energies (eV) for the lowest 5 singlet states for all molecules in Thiel's molecular benchmark set calculated by BSE@$G^0W^0$@PBE with TDA using the aug-cc-pV5Z basis and the FHI-aims-2009 ``Tier2+2aug'' basis set.}
    \begin{tabular}{ c c c c c c c c c c c }
    \hline
    & \multicolumn{5}{c} {aug-cc-pV5Z} & \multicolumn{5}{c} {Tier2+2aug} \\
Molecules & 1 & 2 & 3 & 4 & 5 & 1 & 2 & 3 & 4 & 5 \\
    \hline
	Ethene & 6.24 & 6.64 & 6.93 & 7.27 & 7.66 & 6.20 & 6.57 & 6.83 & 7.37 & 7.75 \\
	E-Butadiene & 5.30 & 5.41 & 5.53 & 5.67 & 5.68 & 5.21 & 5.29 & 5.56 & 5.57 & 5.72 \\
	all-E-Hexatriene & 4.57 & 4.85 & 4.90 & 4.95 & 5.16 & 4.60 & 4.72 & 4.79 & 4.97 & 5.03 \\
	all-E-Octatetraene & 3.92 & 4.25 & 4.53 & 4.63 & 4.90 & 3.98 & 4.28 & 4.44 & 4.56 & 4.80 \\
	Cyclopropene & 5.57 & 6.00 & 6.01 & 6.09 & 6.26 & 5.59 & 5.97 & 6.01 & 6.04 & 6.20 \\
	Cyclopentadiene & 4.65 & 4.80 & 5.40 & 5.41 & 5.53 & 4.68 & 4.69 & 5.22 & 5.23 & 5.44 \\
	Norbornadiene & 4.41 & 5.00 & 5.01 & 5.38 & 5.59 & 4.51 & 5.05 & 5.07 & 5.39 & 5.58 \\
	Benzene & 5.05 & 5.49 & 5.55 & 5.61 & 5.76 & 4.95 & 5.46 & 5.53 & 5.61 & 5.64 \\
	Naphthalene & 4.20 & 4.97 & 4.99 & 5.06 & 5.20 & 4.13 & 4.87 & 4.91 & 4.96 & 5.26 \\
	Furan & 4.67 & 5.18 & 5.36 & 5.47 & 5.71 & 4.57 & 5.24 & 5.36 & 5.43 & 5.62 \\
	Pyrrole & 3.75 & 3.98 & 4.30 & 5.54 & 5.74 & 3.81 & 4.01 & 4.36 & 5.60 & 5.66 \\
	Imidazole & 2.96 & 3.64 & 4.17 & 4.40 & 5.14 & 2.98 & 3.63 & 4.28 & 4.39 & 5.09 \\
	Pyridine & 3.20 & 3.53 & 4.47 & 4.63 & 4.88 & 3.23 & 3.54 & 4.54 & 4.67 & 4.90 \\
	Pyrazine & 2.57 & 3.08 & 4.31 & 4.42 & 4.95 & 2.59 & 3.08 & 4.38 & 4.46 & 4.97 \\
	Pyrimidine & 3.29 & 3.44 & 3.44 & 3.52 & 4.75 & 3.34 & 3.49 & 3.49 & 3.57 & 4.80 \\
	Pyridazine & 1.20 & 2.42 & 3.45 & 3.98 & 4.07 & 1.25 & 2.43 & 3.53 & 4.06 & 4.16 \\
	s-Triazine & 4.22 & 5.40 & 5.57 & 5.57 & 6.11 & 4.25 & 5.45 & 5.45 & 5.45 & 5.92 \\
	s-Tetrazine & 3.42 & 3.67 & 4.62 & 4.84 & 5.00 & 3.47 & 3.75 & 4.66 & 4.78 & 5.05 \\
	Formaldehyde & 5.93 & 6.73 & 6.82 & 7.59 & 7.59 & 5.77 & 6.59 & 6.64 & 7.43 & 7.58 \\
	Acetone & 3.05 & 5.31 & 6.37 & 6.42 & 6.54 & 3.06 & 5.17 & 6.19 & 6.25 & 6.38 \\
	p-Benzoquinone & 1.48 & 1.56 & 3.08 & 4.41 & 4.42 & 1.50 & 1.57 & 3.19 & 4.39 & 4.47 \\
	Formamide & 4.02 & 5.28 & 5.57 & 5.80 & 6.21 & 4.03 & 5.14 & 5.39 & 5.65 & 6.04 \\
	Acetamide & 3.98 & 5.18 & 5.39 & 5.92 & 6.22 & 3.94 & 5.00 & 5.24 & 5.78 & 6.07 \\
	Propanamide & 4.03 & 5.24 & 5.34 & 5.89 & 6.19 & 3.92 & 5.02 & 5.23 & 5.72 & 5.97 \\
	Cytosine & 3.68 & 3.96 & 4.21 & 4.41 & 4.56 & 3.70 & 3.85 & 4.17 & 4.31 & 4.52 \\
	Thymine & 3.58 & 4.15 & 4.76 & 4.87 & 5.04 & 3.61 & 4.22 & 4.76 & 4.83 & 5.08 \\
	Uracil & 3.50 & 4.28 & 4.66 & 4.95 & 5.10 & 3.59 & 4.30 & 4.76 & 4.91 & 4.98 \\
	\hline
    \end{tabular}\\
    \label{tab:bseconv}
\end{table}

\newpage
\section{Basis set convergence of LR-TDDFT in FHI-aims}

\begin{table}[h]
    \centering
	\caption{Neutral excitation energies (eV) for the lowest 5 singlet states for all molecules in Thiel's molecular benchmark set calculated by LR-TDDFT@PBE using the aug-cc-pV5Z basis and the FHI-aims-2009 ``Tier2+aug2'' basis set.}
    \begin{tabular}{ c c c c c c c c c c c }
    \hline
    & \multicolumn{5}{c} {aug-cc-pV5Z} & \multicolumn{5}{c} {Tier2+aug2} \\
Molecules & 1 & 2 & 3 & 4 & 5 & 1 & 2 & 3 & 4 & 5 \\
    \hline
	Ethene & 6.44 & 6.93 & 6.97 & 7.69 & 7.70 & 6.45 & 6.94 & 6.98 & 7.71 & 7.72 \\
	E-Butadiene & 5.52 & 5.76 & 5.77 & 5.81 & 6.13 & 5.54 & 5.76 & 5.77 & 5.82 & 6.14 \\
	all-E-Hexatriene & 4.76 & 5.05 & 5.06 & 5.14 & 5.25 & 4.75 & 5.06 & 5.08 & 5.14 & 5.25 \\
	all-E-Octatetraene & 4.03 & 4.20 & 4.76 & 4.78 & 4.92 & 4.02 & 4.20 & 4.76 & 4.80 & 4.92 \\
	Cyclopropene & 5.83 & 6.14 & 6.17 & 6.28 & 6.35 & 5.85 & 6.15 & 6.18 & 6.29 & 6.34 \\
	Cyclopentadiene & 4.88 & 5.42 & 5.42 & 5.53 & 5.63 & 4.89 & 5.42 & 5.42 & 5.52 & 5.64 \\
	Norbornadiene & 4.57 & 4.92 & 4.98 & 5.42 & 5.46 & 4.58 & 4.92 & 4.99 & 5.42 & 5.46 \\
	Benzene & 3.84 & 4.02 & 4.02 & 4.11 & 6.02 & 3.83 & 4.01 & 4.01 & 4.11 & 6.02 \\
	Naphthalene & 1.92 & 2.91 & 4.19 & 4.66 & 4.82 & 1.91 & 2.90 & 4.18 & 4.66 & 4.81 \\
	Furan & 5.31 & 5.81 & 5.81 & 6.32 & 6.32 & 5.32 & 5.83 & 5.83 & 6.32 & 6.32 \\
	Pyrrole & 4.26 & 4.26 & 4.96 & 4.98 & 5.26 & 4.26 & 4.27 & 4.97 & 4.99 & 5.27 \\
	Imidazole & 5.27 & 5.70 & 5.83 & 6.03 & 6.29 & 5.28 & 5.70 & 5.83 & 6.04 & 6.32 \\
	Pyridine & 4.47 & 5.18 & 5.20 & 5.27 & 5.59 & 4.47 & 5.19 & 5.21 & 5.27 & 5.63 \\
	Pyrazine & 4.86 & 5.51 & 5.65 & 5.83 & 5.88 & 4.86 & 5.51 & 5.66 & 5.84 & 5.88 \\
	Pyrimidine & 4.40 & 4.47 & 5.41 & 5.45 & 5.89 & 4.40 & 4.47 & 5.42 & 5.46 & 5.89 \\
	Pyridazine & 3.60 & 4.07 & 5.14 & 5.42 & 5.46 & 3.60 & 4.06 & 5.14 & 5.42 & 5.47 \\
	s-Triazine & 3.81 & 4.03 & 5.14 & 5.37 & 5.43 & 3.80 & 4.03 & 5.14 & 5.36 & 5.43 \\
	s-Tetrazine & 3.22 & 3.57 & 4.79 & 5.05 & 5.35 & 3.21 & 3.56 & 4.79 & 5.05 & 5.34 \\
	Formaldehyde & 3.81 & 5.81 & 6.58 & 6.77 & 7.08 & 3.81 & 5.82 & 6.57 & 6.78 & 7.10 \\
	Acetone & 4.22 & 4.96 & 5.77 & 5.91 & 6.06 & 4.22 & 4.97 & 5.77 & 5.92 & 6.07 \\
	p-Benzoquinone & 1.91 & 2.05 & 3.48 & 4.35 & 4.45 & 1.91 & 2.05 & 3.48 & 4.35 & 4.45 \\
	Formamide & 5.31 & 5.36 & 5.88 & 5.99 & 6.37 & 5.31 & 5.36 & 5.88 & 5.99 & 6.38 \\
	Acetamide & 4.87 & 5.26 & 5.67 & 5.72 & 5.99 & 4.87 & 5.26 & 5.67 & 5.72 & 6.00 \\
	Propanamide & 4.91 & 5.27 & 5.55 & 5.66 & 5.87 & 4.90 & 5.27 & 5.55 & 5.65 & 5.87 \\
	Cytosine & 3.76 & 4.33 & 4.51 & 4.75 & 4.79 & 3.76 & 4.32 & 4.51 & 4.75 & 4.79 \\
	Thymine & 4.04 & 4.67 & 4.73 & 5.09 & 5.23 & 4.04 & 4.67 & 4.73 & 5.09 & 5.22 \\
	Uracil & 3.93 & 4.70 & 4.83 & 5.17 & 5.18 & 3.93 & 4.70 & 4.82 & 5.17 & 5.18 \\
	Adenine & 6.44 & 6.93 & 6.97 & 7.69 & 7.70 & 6.45 & 6.94 & 6.98 & 7.71 & 7.72 \\
	\hline
    \end{tabular}\\
    \label{tab:tddftconv}
\end{table}

\newpage
\begin{table}[h]
    \centering
	\caption{Neutral excitation energies (eV) for the lowest 5 triplet states for all molecules in Thiel's molecular benchmark set calculated by LR-TDDFT@PBE using the aug-cc-pV5Z basis and the FHI-aims-2009 ``Tier2+aug2'' basis set.}
    \begin{tabular}{ c c c c c c c c c c c }
    \hline
    & \multicolumn{5}{c} {aug-cc-pV5Z} & \multicolumn{5}{c} {Tier2+aug2} \\
Molecules & 1 & 2 & 3 & 4 & 5 & 1 & 2 & 3 & 4 & 5 \\
    \hline
    Ethene & 5.64 & 6.46 & 6.93 & 7.00 & 7.43 & 5.64 & 6.47 & 6.94 & 7.01 & 7.44 \\
	E-Butadiene & 3.92 & 5.54 & 5.78 & 5.82 & 5.99 & 3.92 & 5.56 & 5.78 & 5.82 & 6.01 \\
	all-E-Hexatriene & 3.03 & 4.88 & 5.06 & 5.08 & 5.14 & 3.03 & 4.88 & 5.06 & 5.10 & 5.14 \\
	all-E-Octatetraene & 2.49 & 4.04 & 4.23 & 4.76 & 4.79 & 2.49 & 4.04 & 4.24 & 4.76 & 4.82 \\
	Cyclopropene & 4.93 & 5.87 & 6.02 & 6.16 & 6.16 & 4.93 & 5.89 & 6.03 & 6.16 & 6.16 \\
	Cyclopentadiene & 3.79 & 4.91 & 5.43 & 5.45 & 5.62 & 3.80 & 4.92 & 5.44 & 5.45 & 5.64 \\
	Norbornadiene & 3.92 & 4.70 & 4.93 & 5.18 & 5.42 & 3.93 & 4.70 & 4.93 & 5.19 & 5.42 \\
	Benzene & 3.82 & 3.82 & 3.82 & 3.82 & 5.80 & 3.81 & 3.81 & 3.81 & 3.81 & 5.81 \\
	Naphthalene & 1.57 & 2.77 & 3.78 & 4.33 & 4.54 & 1.56 & 2.76 & 3.77 & 4.33 & 4.54 \\
	Furan & 5.10 & 5.10 & 5.10 & 5.10 & 5.83 & 5.11 & 5.11 & 5.11 & 5.11 & 5.84 \\
	Pyrrole & 3.38 & 4.12 & 4.14 & 4.86 & 4.88 & 3.38 & 4.13 & 4.15 & 4.87 & 4.89 \\
	Imidazole & 4.73 & 5.30 & 5.73 & 5.84 & 5.99 & 4.74 & 5.31 & 5.73 & 5.84 & 6.00 \\
	Pyridine & 4.49 & 5.00 & 5.21 & 5.23 & 5.27 & 4.49 & 5.00 & 5.21 & 5.24 & 5.27 \\
	Pyrazine & 4.88 & 5.13 & 5.47 & 5.68 & 5.73 & 4.88 & 5.14 & 5.47 & 5.68 & 5.73 \\
	Pyrimidine & 4.05 & 4.42 & 4.79 & 5.17 & 5.39 & 4.04 & 4.42 & 4.80 & 5.17 & 5.39 \\
	Pyridazine & 3.24 & 4.04 & 4.43 & 4.72 & 5.23 & 3.24 & 4.03 & 4.43 & 4.71 & 5.23 \\
	s-Triazine & 3.58 & 3.93 & 4.82 & 4.99 & 5.18 & 3.57 & 3.93 & 4.82 & 4.99 & 5.18 \\
	s-Tetrazine & 2.83 & 3.41 & 4.76 & 4.78 & 4.86 & 2.82 & 3.41 & 4.75 & 4.78 & 4.87 \\
	Formaldehyde & 3.48 & 5.83 & 6.59 & 6.78 & 7.12 & 3.48 & 5.84 & 6.60 & 6.79 & 7.14 \\
	Acetone & 3.92 & 4.96 & 5.71 & 5.92 & 6.07 & 3.92 & 4.96 & 5.71 & 5.92 & 6.08 \\
	p-Benzoquinone & 1.70 & 1.85 & 3.06 & 3.17 & 4.25 & 1.70 & 1.85 & 3.06 & 3.17 & 4.25 \\
	Formamide & 5.11 & 5.19 & 5.88 & 5.88 & 5.96 & 5.11 & 5.19 & 5.87 & 5.88 & 5.96 \\
	Acetamide & 4.86 & 5.04 & 5.65 & 5.67 & 5.83 & 4.86 & 5.04 & 5.65 & 5.67 & 5.83 \\
	Propanamide & 4.89 & 5.06 & 5.49 & 5.64 & 5.81 & 4.88 & 5.06 & 5.49 & 5.64 & 5.81 \\
	Cytosine & 3.69 & 3.71 & 4.22 & 4.31 & 4.76 & 3.69 & 3.71 & 4.22 & 4.31 & 4.76 \\
	Thymine & 3.78 & 3.91 & 4.66 & 4.75 & 5.03 & 3.78 & 3.91 & 4.66 & 4.75 & 5.04 \\
	Uracil & 3.81 & 3.96 & 4.63 & 4.68 & 5.06 & 3.81 & 3.96 & 4.63 & 4.68 & 5.06 \\
	Adenine & 6.44 & 6.93 & 6.97 & 7.69 & 7.70 & 6.45 & 6.94 & 6.98 & 7.71 & 7.72 \\
	\hline
    \end{tabular}\\
    \label{tab:tddft_basis_conv}
\end{table}


%